\documentclass[pra,10pt,twocolumn,superscriptaddress,showpacs]{revtex4-1}

\usepackage{epsfig}
\usepackage{amsmath}
\usepackage{graphicx}
\usepackage{graphics}
\usepackage{float}
\usepackage{amssymb}
\usepackage[usenames,dvipsnames]{color}
\usepackage{natbib}
\usepackage{epstopdf}
\usepackage{verbatim}
\usepackage{wrapfig}
\usepackage{mathtools}
\usepackage{dsfont}

\usepackage{breqn}

\graphicspath{{Figure/}}

\usepackage{bm}

\makeatletter
\let\cat@comma@active\@empty
\makeatother

\newcommand{\exs}[1]{\left\langle #1 \right\rangle}

\newcommand{\bra}[1]{\left\langle #1\right|}
\newcommand{\ket}[1]{\left| #1\right\rangle}

\newcommand{\opav}[3]{\langle #1 | #2 | #3 \rangle}

\newcommand{\beq}{\begin{equation}}
\newcommand{\eeq}{\end{equation}}

\begin{document}

\title{Continuous dynamical decoupling of spin chains: modulating the spin-environment and spin-spin interactions}

\author{Sharoon Austin}
\affiliation{School of Science \& Engineering, Lahore University of Management Sciences (LUMS), Opposite Sector U, D.H.A, Lahore 54792, Pakistan}

\author{Muhammad Qasim Khan}
\affiliation{School of Science \& Engineering, Lahore University of Management Sciences (LUMS), Opposite Sector U, D.H.A, Lahore 54792, Pakistan}

\author{Maryam Mudassar}
\affiliation{School of Science \& Engineering, Lahore University of Management Sciences (LUMS), Opposite Sector U, D.H.A, Lahore 54792, Pakistan}

\author{Adam Zaman Chaudhry}
\email{adam.zaman@lums.edu.pk}
\affiliation{School of Science \& Engineering, Lahore University of Management Sciences (LUMS), Opposite Sector U, D.H.A, Lahore 54792, Pakistan}

\begin{abstract}

For spins chains to be useful for quantum information processing tasks, the interaction between the spin chain and its environment generally needs to be suppressed. In this paper, we propose the use of strong static and oscillating control fields in order to effectively remove the spin chain-environment interaction. We find that our control fields can also effectively transform the spin chain Hamiltonian. In particular, interaction terms which are absent in the original spin chain Hamiltonian appear in the time-averaged effective Hamiltonian once the control fields are applied, implying that spin-spin interactions can be engineered via the application of static and oscillating control fields. This transformation of the spin chain can then potentially be used to improve the performance of the spin chain for quantum information processing tasks. For example, our control fields can be used to achieve almost perfect quantum state transfer across a spin chain even in the presence of noise.  As another example, we show how the use of particular static and oscillating control fields not only suppresses the effect of the environment, but can also improve the generation of two-spin entanglement in the spin chain. 

\end{abstract}

\pacs{03.65.Yz, 75.10.Pq, 03.67.Pp, 42.50.Dv}

\maketitle

\section{Introduction}

Spin chains have been a subject of constant study for many years now in diverse areas. For example, spin chains have been used to study phase transitions \cite{Pfeuty1970, PachosPRL2005}, quantum chaos \cite{SantosAmJPhys2012}, high-temperature superconductivity \cite{VuleticPhysRep2006}, and Anderson localization \cite{Anderson1958}. On the experimental front, the physical realization of spin chains ranges from trapped ions \cite{IslamNatCommun2011} to optical lattices \cite{PachosPRL2003}, solid state setups \cite{MajerNature2007}, and photonic systems \cite{GrafeNatPhoton2014}. In the context of quantum information and computation, spin chains have been, for example, extensively studied to achieve perfect quantum state transfer from one site to another \cite{BosePRL2003, KayPRA2006, GodsilPRL2012}, and to generate and distribute entanglement \cite{ClarkNewJPhys2005, SpillerNewJPhys2007, BanchiPRL2011, SahlingNatPhys2015, EstarellasPRA2017}.  However, one of the major hurdles towards the use of spin chains in such quantum information tasks is the inevitable coupling of the spin chain to its environment \cite{BPbook,Weissbook}, which results in the rapid decoherence of the fragile, generally many-body entangled, quantum spin chain state. As such, it is worthwhile studying ways in which the quantum spin chain can be effectively protected from its environment. 

One promising method of protecting the quantum spin chain is to use dynamical decoupling \cite{ViolaPRA1998, LloydPRL1999, UhrigPRL2007,ViolaPRL2010, WestPRL2010, HansonScience2010, LiangJiangPRA2011, WangPRL2011, ChaudhryPRA2014, ManovitzPRL2017, LidarPRL2018}. In dynamical decoupling, control fields are applied rapidly on the quantum system that needs to be protected. The usual approach is to consider different pulse sequences applied to the system  \cite{ViolaPRA1998, Carr1954, UhrigPRL2007, ChaudhryPRA2014, ChaudhryPRA2015} which effectively modulate the system-environment interaction, thereby greatly extending the decoherence timescale. However, one can envisage applying instead strong static and oscillating control fields to dynamically decouple the spin chain, as has been done for a single qubit \cite{FanchiniPRA12007, ChaudhryPRA2013a}, two qubits \cite{FanchiniPRA22007, ChaudhryPRA12012, FanchiniPRA2015}, and an effective large spin system \cite{ChaudhryPRA22012}. This scheme has the advantage that one need not worry about the timing of the different fields; one simply turns on the required fields to achieve effective decoupling of the system from its environment. However, at the same time, the spins in the spin chain are also interacting, and, as a result of the control fields applied, this interaction is also modulated. Consequently, the interactions between the spins are also changed due to the static and oscillating control fields. Instead of considering the change of the spin-spin interactions as a nuisance, we can think about using this change to our advantage. That is, can we apply simple static and oscillating fields to the spin chain such that not only is the spin chain effectively decoupled from its environment (at least to lowest order), but the spin chain Hamiltonian is also changed in such a way that, for example, state transfer fidelity improves or the performance of the spin chain in generating entanglement increases? This is the question that we intend to answer in this paper. We note that control fields, in the form of pulse sequences, have been used to engineer spin chain Hamiltonians \cite{AjoyPRL2013,FrydrychPRA2014,HayesNJP2014,ChoiPRL2017}; however, these pulse sequences can be rather complicated. Our static and oscillating control fields can be used in conjunction with schemes based on pulses, thereby realizing hybrid Hamiltonian engineering techniques. 

We start by considering the Hamiltonian of a general one-dimensional spin chain which is an anisotropic version of the usual XYZ Hamiltonian \cite{Parkinsonbook}. We assume that each spin in this spin chain is coupled `locally' to its environment \cite{FanchiniPRA22007,ChaudhryPRA12012}. In such a situation, we first find suitable static and oscillating control fields that, when applied to the spin chain, are able to dynamically decouple the spin chain, at least to lowest order. The nice feature of these control fields is that the same field needs to be applied to each spin. We then proceed to investigate how the spin chain interactions are modulated by these continuous dynamical decoupling control fields. We find that the spin-spin interactions fundamentally change depending on the control fields, and the effective spin chain Hamiltonian contains interactions that are not present in the original spin chain Hamiltonian. Interestingly, for a special set of control fields, even more additional interaction terms, similar to those in the Dzaolyshinskii-Moriya interaction \cite{Dzaolyshinskiiarticle, Moriya1960, KargarianPRA2009,JafariPRB2008,JafariPRA2014}, can be generated. Our aim then is to analyze the spin chain with these control fields. We first look at the possibility of achieving perfect quantum state transfer by removing the effect of the environment and, at the same time, suitably engineering the spin chain Hamiltonian. We then investigate the entanglement generated between two spins of the spin chain via the spin chain interactions. To this end, we present numerical simulations that first show that the control fields are able to effectively dynamically decouple the spin chain. Second, the simulations show that the dynamics of the spin chain in the presence of the control fields is captured very well by the effective time-averaged Hamiltonian which, in general, contains additional interaction terms. Third, we show that for special control fields, the generation of entanglement can be enhanced even more due to the additional interaction terms. After these numerical simulations, we subsequently endeavor to analytically solve the dynamics of the time-averaged effective spin chain Hamiltonian. We show that that if we impose a condition on the coupling coefficients in the spin chain, we can transform our problem to a system of non-interacting fermions via the Jordan-Wigner transformation \cite{Parkinsonbook}. With this approach, we are able to significantly reduce the computational complexity of the problem. We then demonstrate that our special control fields are able to enhance the entanglement generation, even for larger spin chains. 

This paper is organized as follows. In Sec.~II, we present the static and oscillating control fields we use to dynamically decouple the spin chain from its environment, and derive the effective spin chain Hamiltonian in the presence of these control fields. The use of these control fields towards obtaining perfect quantum state transfer is investigated in Sec.~III. The performance of the control fields in entanglement generation is then numerically analyzed in Sec.~IV. In Sec.~V, we demonstrate results for entanglement generation with relatively larger spin chains, obtained after diagonalizing the effective Hamiltonian via the Jordan-Wigner transformation. Finally, we conclude in Sec.~VI. This is followed by a series of Appendices. In Appendix A, we present the theory behind our dynamical decoupling method and the effective Hamiltonian approach. In Appendix B, we show how effective transverse fields can be included, at least in principle, in the time-averaged Hamiltonian by adding more control fields. Our method of simulating the effect of noise via Ornstein-Uhlenbeck processes is outlined in Appendix C, while Appendix D shows how a single spin operation, such as a spin flip, can be executed, at least in principle, with extremely high fidelity via suitable continuous control fields. Details of the Jordan-Wigner transformation are presented in Appendix E. Finally, in Appendix F, we investigate the degree of fine tuning required in the special control fields in order to generate significant amounts of entanglement.

\section{The formalism}

We start by considering the usual XYZ Hamiltonian which describes a one-dimensional spin chain. Considering only nearest-neighbor coupling, the Hamiltonian, with zero magnetic field, can be written as (we take $\hbar = 1$ throughout)
\begin{equation}
\label{originalspinchainHamiltonian}
H_0 = \sum_{j=1}^{N-1} \sum_{k=1}^{3} \lambda_{jk}\sigma_{k}^{(j)}\sigma_{k}^{(j+1)}.
\end{equation}
Here $\lambda_{jk}$ are the coupling strengths between the spins, $j$ labels the sites, and $k = 1, 2, 3$ denotes $x$, $y$ and $z$ respectively. As usual, $[\sigma_l^{(p)},\sigma_m^{(q)}] = 2i\delta_{pq}\varepsilon_{lmn}\sigma_n^{(p)}$, and note that we are not using cyclic boundary conditions. We want to dynamically decouple the spin chain from its environment. To this end, we first need to model the spin chain-environment interaction. We assume that each spin interacts `locally' with the environment so that the interaction between the spin chain and its environment is given by 
\begin{equation}
\label{systemenvironmentinteraction}
H_{\text{SB}}=\sum_{j=1}^{N-1} B_{x}^{(j)}\sigma_{x}^{(j)}+B_{y}^{(j)}\sigma_{y}^{(j)}+B_{z}^{(j)}\sigma_{z}^{(j)}.
\end{equation}
Here $B_k^{(j)}$ are arbitrary environment operators (or randomly fluctuating noise terms for a classical bath). Our basic strategy is to apply periodic control fields to the spin chain to modulate the interaction between the spin chain and its environment in such a way that the spin chain becomes effectively decoupled from its environment, at least to lowest order. Corresponding to these continuous control fields, there is a unitary operator $U_c(t)$ such that $i \frac{\partial U_c(t)}{\partial t} = H_c(t)U_c(t)$, where $H_c(t)$ is the Hamiltonian describing the action of the control fields on the spin chain. Since we are considering periodic control fields, $U_c(t + t_c) = U_c(t)$. Furthermore, in order to decouple the spin chain from the environment to lowest order, we have the condition \cite{FanchiniPRA12007,ChaudhryPRA12012,ChaudhryPRA2013a,FanchiniPRA2015}
\begin{dmath}
\label{DDcondition}
 \int_0^{t_c}\, dt\, U_c^\dagger(t) H_{\text{SB}} U_c(t) = 0. 
\end{dmath}
For completeness, the reasoning behind this condition is shown in Appendix \ref{eliminatinginteraction}. Keeping the form of $H_{\text{SB}}$ in mind, we guess that 
\begin{dmath}
U_{c}(t)=\prod_{i=1}^{N} e^{i\omega n_{x}\sigma_{x}^{(i)} t}e^{i\omega n_{y}\sigma^{(i)}_{y}t},
\end{dmath}
where $\omega = 2\pi/t_c$ and $n_x$ and $n_y$ are integers, is one possible choice that can dynamically decouple the spin chain from its environment. Our task then is to check that this is indeed the case. It is trivial to check that $U_c(t + t_c) = U_c(t)$. We next define, for convenience, 
\begin{dmath*}
h_{j,k}(t) = U_{c}^{\dagger}(t)\sigma_{k}^{(j)}U_{c}(t),
\end{dmath*}
with $\sigma_1^{(j)} = \sigma_x^{(j)}$, $\sigma_2^{(j)} = \sigma_y^{(j)}$, and $\sigma_3^{(j)} = \sigma_z^{(j)}$. We find that  
\begin{align*}
h_{j,1}(t) &=\cos(2\omega n_{y}t)\sigma_{x}^{(j)}-\sin(2\omega n_{y}t)\sigma_{z}^{(j)}, \\
h_{j,2}(t) &=\sin(2\omega n_{x}t)\sin(2\omega n_{y}t)\sigma_{x}^{(j)} +\\
&\cos(2\omega n_{x}t)\sigma_{y}^{(j)}+\sin(2\omega n_{x}t)\cos(2\omega n_{y}t)\sigma_{z}^{(j)}, \\
h_{j,3}(t) &=\cos(2\omega n_{x}t)\sin(2\omega n_{y}t)\sigma_{x}^{(j)}-\\
&\sin(2\omega n_{x}t)\sigma_{y}^{(j)}+ \cos(2\omega n_{x}t)\cos(2\omega n_{y}t)\sigma_{z}^{(j)}.
\end{align*}
With these expressions, it is straightforward to see that as long as $n_x \neq n_y$, we meet the condition given by Eq.~\eqref{DDcondition}. The corresponding control field Hamiltonian is  
\begin{dmath}
\label{controlHamiltonian}
H_{c}(t)=\sum_{i=1}^{N}\left\lbrace \omega n_y [\sin(2 \omega n_x t )\sigma_{z}^{(i)} - \cos(2  \omega n_x t)\sigma_{y}^{(i)}] -\omega n_x \sigma_{x}^{(i)} \right\rbrace,
\end{dmath}
with $n_x \neq n_y$. We emphasize that our decoupling scheme works provided that $t_c \ll \tau$, where $\tau$ is the environment correlation time, with exact decoupling achieved in the limit $\frac{t_c}{\tau} \rightarrow 0$ (see Appendix \ref{eliminatinginteraction} for more details). In other words, provided that $\omega$ is large enough, we are able to dynamically decouple the spin chain from its environment, at least to lowest order, by using two oscillating fields in the $y$ and $z$ directions and a static field in the $x$ direction. Then, as long as each spin interacts `locally' with the environment [see Eq.~\eqref{systemenvironmentinteraction}], independent of the detailed form of the spin chain-environment interaction, the spin chain can be effectively decoupled from the environment. 

We now observe that the control fields not only serve to dynamically decouple the spin chain, but they also modify the spin chain Hamiltonian itself. Provided that the control fields are strong enough and oscillating fast enough, the effective spin chain Hamiltonian in the presence of the control fields is  \cite{ChaudhryPRA12012,ChaudhryPRA2013a}
\begin{dmath*}
\bar{H} = \frac{1}{t_c} \int_0^{t_c} \, dt \, U_c^\dagger (t) H_0 U_c(t).
\end{dmath*}
For completeness, this relation is also derived in Appendix A. In particular, our effective Hamiltonian approach is valid if $\lambda_{j,k}t_c \ll 1$, with $\bar{H}$ able to capture the dynamics perfectly in the limit $\lambda_{j,k}t_c \rightarrow 0$. In our case, the effective Hamiltonian becomes  
\begin{dmath*}
\bar{H}=\frac{1}{t_{c}}\sum_{j=1}^{N-1} \int_{0}^{t_c} \, dt \, \sum_{k=1}^{3} \lambda_{jk} h_{j,k}(t)h_{j+1,k}(t).
\end{dmath*}
We now define 
\begin{align*}
I_1^{(j)} &=\frac{1}{t_c}\int_0^{t_c} h_{j,1}(t) h_{j+1,1}(t)\, dt, \\
I_2^{(j)} &=\frac{1}{t_c}\int_0^{t_c} h_{j,2}(t) h_{j+1,2}(t) \, dt, \\
I_3^{(j)}&=\frac{1}{t_c}\int_0^{t_c} h_{j,3}(t) h_{j+1,3}(t) \, dt.
\end{align*}
The effective Hamiltonian is then 
\begin{equation*}
\bar{H} = \sum_{j=1}^{N-1} \left[\lambda_{j1}I_1^{(j)} +\lambda_{j2} I_2^{(j)} + \lambda_{j3} I_3^{(j)}\right].
\end{equation*}
The remaining task is to evaluate the integrals. Recalling that $n_x$ and $n_y$ are integers with $n_x \neq n_y$ (since we want to dynamically decouple the spin chain from its environment), we find that, if $n_y \neq 2n_x$,
\begin{align*}
I_1^{(j)} &= \frac{1}{2}[\sigma_{x}^{(j)}\sigma_{x}^{(j+1)}  +  \sigma_{z}^{(j)}\sigma_{z}^{(j+1)}], \\
I_2^{(j)}&=\frac{1}{4}[\sigma_{x}^{(j)}\sigma_{x}^{(j+1)}+2\sigma_{y}^{(j)}\sigma_{y}^{(j+1)}+\sigma_{z}^{(j)}\sigma_{z}^{(j+1)}], 
\end{align*}
and $I_3^{(j)} = I_2^{(j)}$. This leads to 
\begin{dmath}
\bar{H}_1=\sum_{j=1}^{N-1}\left\lbrace \frac{\lambda_{j1}}{2}\left[\sigma_{x}^{(j)}\sigma_{x}^{(j+1)}+\sigma_{z}^{(j)}\sigma_{z}^{(j+1)}\right]+ (\lambda_{j2}+\lambda_{j3})\left[\frac{1}{4}(\sigma_{x}^{(j)}\sigma_{x}^{(j+1)}+2\sigma_{y}^{(j)}\sigma_{y}^{(j+1)}+\sigma_{z}^{(j)}\sigma_{z}^{(j+1)})\right]\right\rbrace,
\label{Hbar1eq}
\end{dmath}
for $n_x \neq n_y$ and $n_y \neq 2n_x$. Thus, by applying local control fields, the spin chain is dynamically decoupled from its environment, and the interactions between the spins are also transformed. While the transformed spin chain Hamiltonian may be more complicated than the original spin chain Hamiltonian, this may not always be the case. For example, one can check that the fully isotropic Heisenberg Hamiltonian, also known as the Heisenberg XXX model, remains unchanged. Also, the modified Hamiltonian may itself be a very well-known and understood model - for instance, the quantum Ising model transforms to the XX model (also known as the isotropic XY model). Even if the effective spin chain Hamiltonian is relatively complicated, it is still tractable for small spin chains; moreover, the effective Hamiltonian can also be studied for larger spin chains in some special cases (see Section V). Throughout the paper, our focus will be on showing how the modified interactions can improve quantum state transfer and entanglement generation.

We now notice that if we use control fields such that $n_y = 2n_x$, $I_1^{(j)}$ is the same as before, but now 
\begin{align*}
I_2^{(j)}&=\frac{1}{4}[\sigma_{x}^{(j)}\sigma_{x}^{(j+1)}+2\sigma_{y}^{(j)}\sigma_{y}^{(j+1)}+\sigma_{x}^{(j)}\sigma_{y}^{(j+1)}\\ &+\sigma_{y}^{(j)}\sigma_{x}^{(j+1)}+\sigma_{z}^{(j)}\sigma_{z}^{(j+1)}], \\
I_3^{(j)}&=\frac{1}{4}[\sigma_{x}^{(j)}\sigma_{x}^{(j+1)}+2\sigma_{y}^{(j)}\sigma_{y}^{(j+1)}-\sigma_{x}^{(j)}\sigma_{y}^{(j+1)} \\ &-\sigma_{y}^{(j)}\sigma_{x}^{(j+1)}+\sigma_{z}^{(j)}\sigma_{z}^{(j+1)}].
\end{align*}
In this case, we can then write the effective Hamiltonian as 
\begin{dmath}
\bar{H}_2 = \sum_{j=1}^{N-1}\frac{\lambda_{j1}}{2}\left[\sigma_{x}^{(j)}\sigma_{x}^{(j+1)}+\sigma_{z}^{(j)}\sigma_{z}^{(j+1)}\right]+ \frac{\lambda_{j2}}{4}\left[\sigma_{x}^{(j)}\sigma_{x}^{(j+1)}+2\sigma_{y}^{(j)}\sigma_{y}^{(j+1)}+\sigma_{x}^{(j)}\sigma_{y}^{(j+1)}+\sigma_{y}^{(j)}\sigma_{x}^{(j+1)}+\sigma_{z}^{(j)}\sigma_{z}^{(j+1)}\right]+\frac{\lambda_{j3}}{4}\left[\sigma_{x}^{(j)}\sigma_{x}^{(j+1)}+2\sigma_{y}^{(j)}\sigma_{y}^{(j+1)}-\sigma_{x}^{(j)}\sigma_{y}^{(j+1)}-\sigma_{y}^{(j)}\sigma_{x}^{(j+1)}+\sigma_{z}^{(j)}\sigma_{z}^{(j+1)}\right].
\label{Hbar2eq}
\end{dmath}
This case is even more interesting due to the additional presence of the `cross-interactions' such as $\sigma_x^{(j)} \sigma_y^{(j + 1)}$. Such `cross-interactions' arise in spin chains when one studies Dzyaloshinskii-Moriya interactions in spin chains (although the signs of our additional terms differ). However, in our case, these interactions are simply an effective result of applying control fields to each spin. As we will show, these additional interactions can significantly improve entanglement generation. 

Let us now also note that the spin chain Hamiltonian that we started from [see Eq.~\eqref{originalspinchainHamiltonian}] does not contain any transverse fields which would contribute $\sum_{j = 1}^N b_{j1} \sigma_x^{(j)} + b_{j2} \sigma_y^{(j)} + b_{j3} \sigma_z^{(j)}$ to the Hamiltonian. This is simply because our dynamical decoupling fields, at least to lowest order, remove the effect of these terms, provided that the time-dependence of the fields $b_{j1}$, $b_{j2}$, and $b_{j3}$ is slow compared to $t_c$. However, if these additional transverse fields are also oscillating with frequency comparable to $\omega$, then the effective Hamiltonian can, at least in principle, include the effect of static fields as well. Further details are presented in Appendix \ref{includingstaticfields} [in particular, see Eq.~\eqref{additionalfieldsforstatic} for the additional control fields that lead to additional terms in the effective Hamiltonian given by Eq.~\eqref{effectivewithstatic1} or Eq.~\eqref{effectivewithstatic2}].

\section{Quantum state transfer}

As a first example of our formalism, we study the transfer of a quantum state from one end of a quantum spin chain to the other \cite{BosePRL2003,NikoJCM2004,ZanardiPRA2007,FrancoPRL2008,Kaytransferreview,RabitzPRA2011,Jexstatetransfer,HorodeckiPRA2014}. The most commonly studied scenario involves the quantum XX model
$$ H_{\text{XX}} =  \sum_{j=1}^{N-1}\lambda_j \left[  \sigma_{x}^{(j)}\sigma_{x}^{(j+1)} +  \sigma_y^{(j)} \sigma_y^{(j+1)}\right] + \sum_{j = 1}^N B_j \sigma_z^{(j)}. $$
The idea is that an arbitrary state for the spin chain can be transferred to the other end. Writing the eigenstates of the $\sigma_z$ operator as $\ket{0}$ and $\ket{1}$ with $\sigma_z \ket{s} = (-1)^s \ket{s}$, it has been found that if the initial state of the spin chain is $\ket{\Psi(0)} = (a\ket{0} + b\ket{1}) \otimes \ket{0} \otimes \hdots \otimes \ket{0}$, the state of the spin chain after some time $T$ is $\ket{\Psi(T)} = \ket{0} \otimes \hdots \otimes \ket{0} \otimes (a\ket{0} + e^{i\phi} b\ket{1})$, where $\phi$ is known so that the phase can be corrected at the end of the state transfer. One way to achieve perfect state transfer is that we set $\lambda_j = \sqrt{j(N - j)}$ \cite{ChristandlPRL2004, NikolopoulosEurophyslett2004}, which is optimal in terms of the transfer time \cite{YungPRA2006}. Perfect state transfer is then achieved after time $T = \pi/2$, with the phase factor $e^{i\phi} = (-i)^{N - 1}$ that can be removed \cite{PetrosyanPRA2010}. Practically speaking, however, such perfect state transfer is difficult due to the unwanted influences of the environment. To remove this detrimental effect, pulse sequences \cite{FrydrychJPhysB2014} have been considered and the direct modulation of the spin-spin coupling has also been investigated \cite{ZwickNJP2014}. With our scheme, as we have discussed, local noise terms can be eliminated to lowest order by applying a static as well as oscillating control fields. These control fields also modify the spin chain Hamiltonian. In particular, it is clear that the quantum XX spin chain does not remain the quantum XX spin chain in the presence of the control fields. To get around this, we note that if we originally have the quantum Ising model (with zero magnetic field),
\begin{equation}
\label{Isingeq}
H_0=\sum_{j=1}^{N-1}\lambda_{j}\sigma_{x}^{(j)}\sigma_{x}^{(j+1)},
\end{equation}
the corresponding time-averaged effective Hamiltonian in the presence of the control fields is 
\begin{align}
\label{effectiveIsing}
\bar{H}_1=\sum_{j=1}^{N-1}\frac{\lambda_{j}}{2}\left[\sigma_{x}^{(j)}\sigma_{x}^{(j+1)}+\sigma_{z}^{(j)}\sigma_{z}^{(j+1)}\right].
\end{align}
In this case, it turns out that $\bar{H}_2 = \bar{H}_1 = \bar{H}$, so $n_y = 2n_x$ and $n_y \neq 2n_x$ lead to the same result. Notice that the effective Hamiltonian contains $\sigma_{z}^{(j)}\sigma_{z}^{(j+1)}$ interactions which are absent in the original Ising chain. In particular, the effective Hamiltonian is simply a rotated version of the XX model. Defining the rotation operator $U_R = \Pi_j e^{-i\pi \sigma_x^{(j)}/4}$, we find that $\bar{H}_R = U_R \bar{H}_1 U_R^\dagger = H_{\text{XX}}$ with $B_j = 0$. In other words, if we start from the Ising spin chain, and apply the control fields given by
\begin{dmath}
\label{controlfieldsising}
H_{c}^R(t)=\sum_{i=1}^{N}\left\lbrace \omega n_z [\sin(2 \omega n_x t )\sigma_{y}^{(i)} + \cos(2  \omega n_x t)\sigma_{z}^{(i)}] -\omega n_x \sigma_{x}^{(i)} \right\rbrace,
\end{dmath}
where $n_x$ and $n_z$ are integers with $n_x \neq n_z$, we can achieve excellent state transfer $(a\ket{0} + b\ket{1}) \otimes \ket{0} \otimes \hdots \otimes \ket{0} \rightarrow \ket{0} \otimes \hdots \otimes \ket{0} \otimes (a\ket{0} + b\ket{1})$ even in the presence of the local noise terms. 

\begin{figure}[h!]
\centering
\includegraphics[width=.8\linewidth]{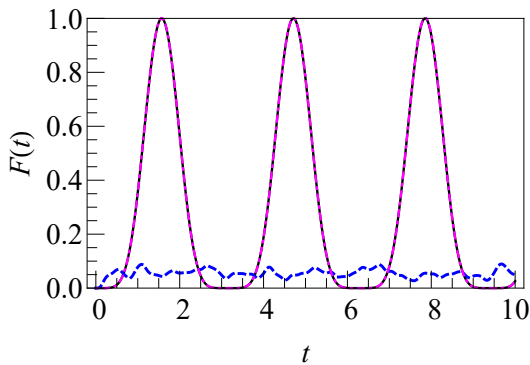}
\caption{(Color online) Dynamics of the fidelity of state transfer starting from the quantum Ising model with $N = 4$. We have used $\lambda_{j}=\sqrt{j(N - j)}$, and $n_{y}=2$, $n_{x} = 1$; the initial state is $\ket{1000}$. The dot-dashed, magenta curve shows the fidelity with the full Hamiltonian $H_{0}+H_{\text{SB}}+H_{c}^R(t)$ numerically solved [$H_0$ and $H_c^R(t)$ are given by Eqs.~\eqref{Isingeq} and \eqref{controlfieldsising} respectively], while the solid, black curve shows the dynamics using the time-averaged effective Hamiltonian $\bar{H}_R$. The dashed blue curve shows the fidelity if we simply use only the XX spin chain $H_{\text{XX}}$ in the presence of noise. Throughout the paper, we work in dimensionless units with $\hbar$ set equal to one and $t_c = 0.01$. Also, as explained in the main text, the noise is modeled via Ornstein-Uhlenbeck processes with zero mean, correlation time $\tau = 0.5$, and standard deviation $\sigma = 2.0$. We take an average over $20$ noise realizations throughout the paper. In liquid NMR implementations of spin chains \cite{Jexstatetransfer}, $\lambda_j$ are typically of the order of $100$ Hz. $t_c = 0.01$ then leads to control fields with frequencies in the $10$ kHz regime and amplitudes around $1$ microtesla, and $t = 1$ is on the order of $10$ ms. With superconducting qubits \cite{MajerNature2007}, $\lambda_j$ is on the order of $10$ MHz, meaning that $t_c = 0.01$ corresponds to control field frequencies in the GHz regime and amplitudes around $0.01$ tesla. Smaller values of $t_c$ correspond to higher frequencies and stronger fields, thereby leading to even better decoupling of the spin chain and the effective Hamiltonian approximating the exact dynamics even more closely.}
\label{statetransferN4}
\end{figure}

\begin{figure}[h!]
\centering
\includegraphics[width=.8\linewidth]{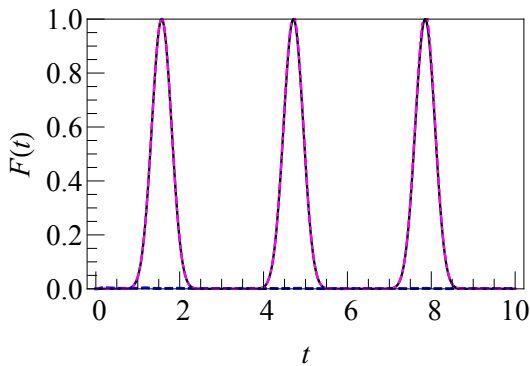}
\caption{(Color online) Same as Fig.~\ref{statetransferN4}, except that we now have $N = 10$. The initial state is $\ket{100\hdots 0}$. The dashed-blue line essentially overlaps with the horizontal axis - the fidelity without control fields is very small.}
\label{statetransferN10}
\end{figure}

We now numerically check our claims. Since the Hamiltonian $\bar{H}_R$ preserves the number of excitations, we restrict ourselves to studying the transfer $\ket{\psi_i} = \ket{1} \otimes \ket{0} \otimes \hdots \otimes \ket{0} \rightarrow \ket{\psi_f} = \ket{0} \otimes \hdots \otimes \ket{1}$ for simplicity. To quantify the quality of the state transfer, we use the fidelity $F(t) = \opav{\psi_f}{\rho(t)}{\psi_j}$ where $\rho(t)$ is the spin state density matrix at time $t$. We model the effect of the environment on the spins via classical noise fields acting on each spin. For simplicity, we assume that $B_k^{(j)}$ is the same for every $j$. $B_x^{(j)}$, $B_y^{(j)}$, and $B_z^{(j)}$ are then generated via independent Ornstein-Uhlenbeck processes, each with zero mean, correlation time $\tau = 0.5$, and standard deviation $\sigma = 2.0$ (for more details, see Appendix \ref{appendixnoise} and Ref.~\cite{Jacobsbook}). Let us also note that our spin state transfer depends on the reliable implementation of a single spin operation in order to prepare the initial state $\ket{1} \otimes \ket{0} \otimes \hdots \otimes \ket{0}$. As illustrated in Appendix \ref{appendixsinglespingate}, our dynamical decoupling control fields can, at least in principle, be extended in order to implement single spin operations with fidelity very close to one.

In Fig.~\ref{statetransferN4}, we illustrate state transfer for $N = 4$. First, the fidelity is captured very well by the effective Hamiltonian since the solid black curve essentially overlaps with the dot-dashed magenta curve. Second, the effect of the noise is effectively removed. Third, if we use the XX model with no control fields and noise present, the fidelity of the state transfer is significantly lower as shown by the dashed blue curve. Similar results are obtained for larger spin chains as illustrated in Fig.~\ref{statetransferN10}. Since the effective Hamiltonian is the XX model, and it is known that (with a proper choice of the spin-spin coupling strengths) the XX model leads to perfect state transfer \cite{Kaytransferreview}, our proposed dynamical decoupling fields should lead to near perfect state transfer for even larger spin chains. Thus, we have demonstrated that if we use the quantum Ising model to begin with, we can obtain excellent quantum state transfer simply by the use of static and oscillating control fields even in the presence of noise.

\section{Numerical results for entanglement generation}
We now quantitatively analyze the results of applying the control fields for entanglement generation. To do this, we look at the concurrence \cite{WoottersPRL1998} between two spins in the spin chain. Our strategy is simple. We consider the spin chain in the presence of local noise fields. To begin, we do not apply any control fields, and examine the behavior of the concurrence between two spins as a function of time. Thereafter, we apply our strong and rapidly oscillating control fields. We find the concurrence between two spins in the presence of the noise fields by solving the Schrodinger equation, thereby showing the effectiveness of the control fields in dynamically decoupling the spin chain. We also show that the dynamical behavior is captured very well by the time-averaged effective Hamiltonian approach. Finally, we compare the performance of the control fields with $n_y \neq 2n_x$ and $n_y = 2n_x$. The initial state we consider is either the fully polarized spin state, that is, $\ket{00\hdots 0}$ (or $\ket{11\hdots 1}$), or the fully polarized state with the first spin flipped, that is, $\ket{100\hdots 0}$ (or $\ket{011\hdots 1}$). The fully polarized spin state can be realized experimentally by, for instance, applying a large magnetic field at low temperatures (the coordinate system is set up such that the $z$-axis is aligned along the magnetic field), and a $\pi$-pulse can be applied to the first spin to realize the spin flip. Indeed, these states are the initial states most commonly used in studies of entanglement dynamics (see, for example, Refs.~\cite{GalvePRA2009} and \cite{WangPRA2001}).  

\begin{figure}[h!]
\centering
\includegraphics[width=.8\linewidth]{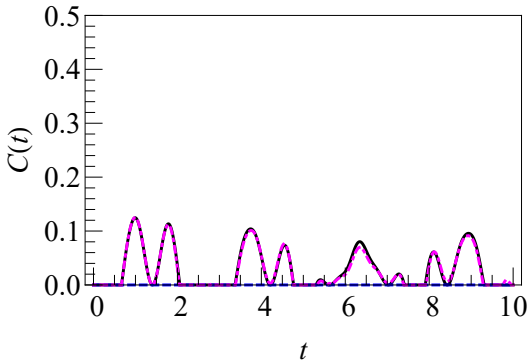}
\caption{(Color online) Dynamics of the concurrence between spins 1 and 4 starting from the quantum Ising model with $N = 4$. We have used $\lambda_{1}=2$ ($\lambda_2 = \lambda_3 = 0$), and $n_{y}=2$, $n_{x} = 1$. The dynamics for the concurrence without any control fields is shown by the dashed, blue curve, while the dot-dashed, magenta curve shows the dynamics with the full Hamiltonian $H_{0}+H_{\text{SB}}+H_{c}(t)$ numerically solved [see Eqs.~\eqref{Isingeq} and \eqref{controlHamiltonian} for $H_0$ and $H_c(t)$ respectively]. The solid, black curve shows the dynamics using the time-averaged effective Hamiltonian. The initial state of the spin chain used here is $\ket{0000}$. As before, we are working in dimensionless units with $\hbar$ set equal to one and $t_c = 0.01$. The dashed, blue line overlaps with the horizontal axis.}
\label{quantumIsing}
\end{figure}

\subsection{The quantum Ising model}
In this case, the spin chain Hamiltonian (with zero magnetic field) is given by Eq.~\eqref{Isingeq}, while the corresponding effective Hamiltonian in the presence of the control fields is shown in Eq.~\eqref{effectiveIsing}. From now on, for simplicity, we will be assuming that the coupling strengths are the same throughout the chain, that is, $\lambda_{jk}$ is independent of $j$. As outlined before, we aim to find the concurrence between two spins in the spin chain as a function of time. We find the density matrix as a function of time, and then take the partial trace over all the spins other than the two spins whose concurrence we are interested in. Having found this two-spin density matrix $\rho_2(t)$ as a function of time, we find the concurrence \cite{WoottersPRL1998} by first finding 
\begin{dmath*}
R=\sqrt{\sqrt{\rho_2} {\widetilde{\rho_2}} \sqrt{\rho_2}},
\end{dmath*}
with 
\begin{dmath*}
\widetilde{\rho}_2=(\sigma_{y} \otimes \sigma_{y})\rho_2^{*} (\sigma_{y} \otimes \sigma_{y}).
\end{dmath*}
The concurrence $C$ is then given by 
\begin{dmath*}
C(\rho)=\max(0,\lambda_{1}-\lambda_{2}-\lambda_{3}-\lambda_{4})
\end{dmath*}
where $\lambda_i$ are the eigenvalues of $R$ in descending order.

\begin{figure}[h!]
\centering
\includegraphics[width=.8\linewidth]{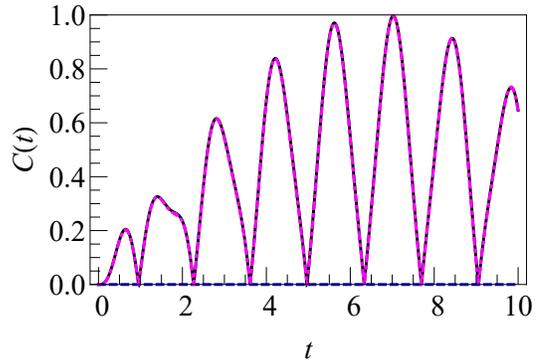}
\caption{(Color online) Same as Fig.~\ref{quantumIsing}, but now we have the control fields given by Eq.~\eqref{controlfieldsising}. That is, the dot-dashed magenta curve shows the dynamics with the Hamiltonian $H_0 + H_{\text{SB}} + H_c^R(t)$, the dynamics without any control fields is shown by the dashed-blue curve, while the solid, black curve shows the dynamics with the effective Hamiltonian. The initial state is $\ket{1000}$. We have checked that when the concurrence is approximately one, the purity of the state for spins $1$ and $4$ is very close to one with the two-spin state being approximately the fully entangled state $\frac{1}{\sqrt{2}}\left(\ket{01} + i\ket{10}\right)$. Once again, the dashed, blue line overlaps with the horizontal axis.}
\label{quantumIsingdifffields}
\end{figure}

Let us now present our results for the quantum Ising model. In Fig.~\ref{quantumIsing}, we illustrate three points. First, as can be seen by comparing the solid, black curve with the dot-dashed, magenta curve, the time-averaged Hamiltonian reproduces the exact numerical results very well. Second, the concurrence in the presence of the noise is very low - the dashed, blue curve overlaps with the horizontal axis. Third, the control fields are able to average out the effect of the noise fields. Given the close relation between entanglement and quantum state transfer \cite{Kaytransferreview}, we have also found the entanglement with the initial state $\ket{1000}$ and the control fields given by Eq.~\eqref{controlfieldsising} [as in Fig.~\ref{statetransferN4}]. The results are shown in Fig.~\ref{quantumIsingdifffields}. We can see that we can not only protect the spin chain against the environment, but also generate almost perfect entanglement between the ends of the spin chain, at least for $N = 4$.  

As we have seen, if we start from the quantum Ising model, there is no difference between $\bar{H}_1$ and $\bar{H}_2$. In order to investigate how the condition $n_y = 2n_x$ can make a difference, we now look at the XY model.

\subsection{XY model}
For the XY model, the spin chain Hamiltonian, with zero magnetic field, is 
\begin{equation*}
H_0=\sum_{j=1}^{N-1} \left[ \lambda_{1} \sigma_{x}^{(j)}\sigma_{x}^{(j+1)} + \lambda_2 \sigma_y^{(j)} \sigma_y^{(j+1)}\right].
\end{equation*}
If $\lambda_1 = \lambda_2$, we have the isotropic XY model, which we have referred to as the XX model. In the presence of the control fields, the effective Hamiltonian is 
\begin{dmath}
\bar{H}_1=\sum_{j=1}^{N-1}\left\lbrace \frac{\lambda_{1}}{2}\left[\sigma_{x}^{(j)}\sigma_{x}^{(j+1)}+\sigma_{z}^{(j)}\sigma_{z}^{(j+1)}\right]+ \frac{\lambda_{2}}{4}\left[\sigma_{x}^{(j)}\sigma_{x}^{(j+1)}+2\sigma_{y}^{(j)}\sigma_{y}^{(j+1)}+\sigma_{z}^{(j)}\sigma_{z}^{(j+1)}\right]\right\rbrace,
\end{dmath}
if $n_y \neq 2n_x$. Once again, note the presence of the additional spin-spin interactions. On the other hand, if $n_y = 2n_x$, the effective Hamiltonian is 
\begin{dmath}
\bar{H}_2 = \sum_{j=1}^{N-1}\left\lbrace\frac{\lambda_1}{2}\left[\sigma_{x}^{(j)}\sigma_{x}^{(j+1)}+\sigma_{z}^{(j)}\sigma_{z}^{(j+1)}\right]+ \frac{\lambda_2}{4}\left[\sigma_{x}^{(j)}\sigma_{x}^{(j+1)}+2\sigma_{y}^{(j)}\sigma_{y}^{(j+1)}+\sigma_{x}^{(j)}\sigma_{y}^{(j+1)}+\sigma_{y}^{(j)}\sigma_{x}^{(j+1)}+\sigma_{z}^{(j)}\sigma_{z}^{(j+1)}\right]\right\rbrace.
\end{dmath}
There are now even more additional spin-spin interactions; $\bar{H}_2$ contains `cross-interaction' terms $\sigma_{x}^{(j)}\sigma_{y}^{(j+1)}$ and $\sigma_{y}^{(j)}\sigma_{x}^{(j+1)}$ absent in $\bar{H}_1$. 

\begin{figure}[h!]
\centering
\includegraphics[width=.8\linewidth]{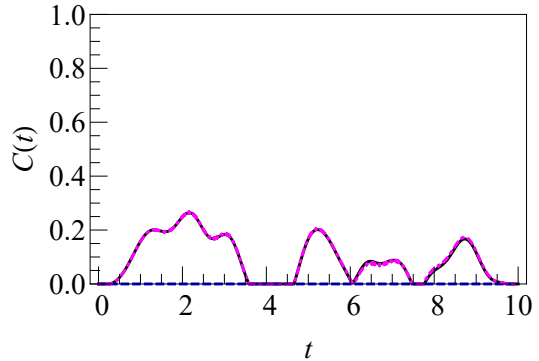}
\caption{(Color online) Plot of the concurrence between spins 1 and 4 starting from the quantum XX model with $N = 4$. Here we have $n_{y} \neq 2n_{x}$ ($n_y = 3, n_x = 1$), and $\lambda_1=\lambda_2 =1$, while $\lambda_3 = 0$. We have used $t_c = 0.01$. We have shown the dynamics without any control fields (dashed, blue curve which is overlapping with the horizontal axis), with control fields using the total Hamiltonian $H_{0}+H_{\text{SB}}+H_{c}(t)$ (dot-dashed, magenta curve), and using the time-averaged Hamitonian $\bar{H}_1$ (solid, black curve). The initial state of the spin chain is $\ket{0000}$. No entanglement can be generated even in the absence of noise since the state $\ket{0000}$ is an eigenstate of the XX model; the control fields not only eliminate the noise, but also modify the spin chain Hamiltonian such that entanglement can be generated.}
\label{Hbar1XYparticles1and4}
\end{figure}

\begin{figure}[h!]
\centering
\includegraphics[width=.8\linewidth]{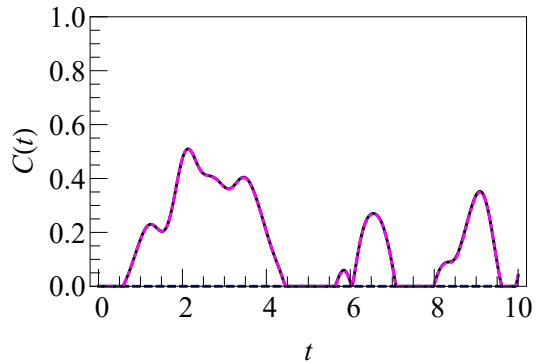}
\caption{(Color online) Same as Fig.~\ref{Hbar1XYparticles1and4}, except that we are now using control fields with $n_y = 2n_x$ ($n_y = 2, n_x = 1$), and the solid, black curve shows the entanglement dynamics due to the effective Hamiltonian $\bar{H}_2$. The initial state is $\ket{0000}$.}
\label{Hbar2XYparticles1and4}
\end{figure}

\begin{figure}[h!]
\centering
\includegraphics[width=.8\linewidth]{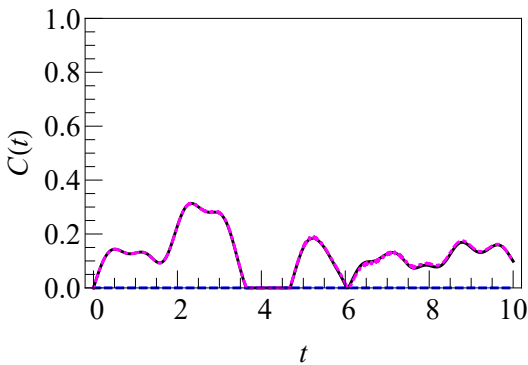}
\caption{(Color online) Same as Fig.~\ref{Hbar1XYparticles1and4}, except that we are now showing the concurrence between spins 2 and 3 of the spin chain. Once again, the initial state is $\ket{0000}$.}
\label{Hbar1XYparticles2and3}
\end{figure}

\begin{figure}[h!]
\centering
\includegraphics[width=.8\linewidth]{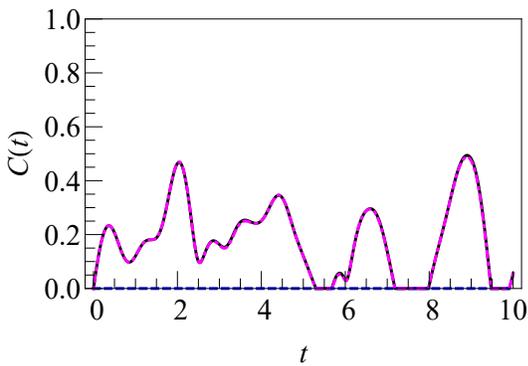}
\caption{(Color online) Same as Fig.~\ref{Hbar1XYparticles1and4}, except that we are now using control fields with $n_y = 2n_x$ ($n_y = 2, n_x = 1$), and we have shown the concurrence between spins 2 and 3 of the spin chain. The solid, black curve shows the entanglement dynamics due to the effective Hamiltonian $\bar{H}_2$. The initial state is again $\ket{0000}$.}
\label{Hbar2XYparticles2and3}
\end{figure}

\begin{figure}[h!]
\centering
\includegraphics[width=.8\linewidth]{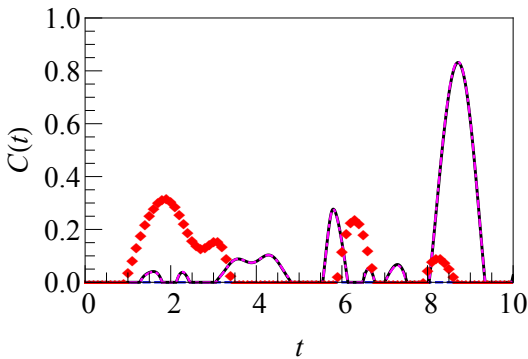}
\caption{(Color online) Plot of the concurrence between spins 1 and 4 for the quantum XY model with $N = 4$. Here we have $n_{y} = 2n_{x}$ ($n_y = 2, n_x = 1$), and $\lambda_1=\lambda_2 =1$, while $\lambda_3 = 0$. We have shown the dynamics without any control fields (solid, dashed blue line), with control fields using the total Hamiltonian $H_{0}+H_{\text{SB}}+H_{c}(t)$ (dot-dashed, magenta curve), and using the time-averaged Hamitonians $\bar{H}_1$ (red diamonds) and $\bar{H}_2$ (solid, black curve). The initial state is now $\ket{0111}$.}
\label{HbarXYp14updownlabel}
\end{figure}

We now present numerical simulations illustrating the effect of these additional terms. First, Fig.~\ref{Hbar1XYparticles1and4} shows the concurrence between the first and last spins of the spin chain with $N = 4$, starting from the initial state $\ket{0000}$. The dashed, blue curve (which is on top of the horizontal axis) illustrates that, in the absence of the control fields, entanglement generation is negligible. However, in the presence of the control fields with $n_y \neq 2n_x$, significant entanglement can be generated, as evidenced by the dot-dashed, magenta curve. Moreover, the solid black curve, which lies essentially on top of the magenta curve, shows that the dynamics are captured very well by the time-averaged effective Hamiltonian $\bar{H}_1$. Moving on, in Fig.~\ref{Hbar2XYparticles1and4}, we have again shown the dynamics of the entanglement between spins $1$ and $4$, but we have now used the special control fields with $n_y = 2n_x$. The initial state is again $\ket{0000}$. As before, considerable entanglement is generated in the presence of the control fields, and the dynamics is captured very well by the effective Hamiltonian (which is $\bar{H}_2$ now). Moreover, comparing Figures \ref{Hbar1XYparticles1and4} and \ref{Hbar2XYparticles1and4}, we see that the special choice $n_y = 2n_x$ is able to generate more entanglement (at least between spins 1 and 4). If we look instead at spins 2 and 3 of the spin chain [see Figures \ref{Hbar1XYparticles2and3} and \ref{Hbar2XYparticles2and3}], we reach a similar conclusion. Thus, while fields with any $n_y$ and $n_x$ can decouple the spin chain (as long as $n_y \neq n_x$), making the special choice $n_y = 2n_x$ can be a much better strategy in the sense that the  generation of a valuable quantum resource such as quantum entanglement is improved. This result is further reinforced in Fig.~\ref{HbarXYp14updownlabel} where we have used a different initial state, namely $\ket{0111}$. Once again, $\bar{H}_2$ can generate significantly more entanglement between the ends of the spin chain as compared to $\bar{H}_1$. Similar conclusions hold true if we use the initial states $\ket{1000}$ and $\ket{1111}$.

\subsection{XYZ model}
\begin{figure}[h!]
\centering
\includegraphics[width=.8\linewidth]{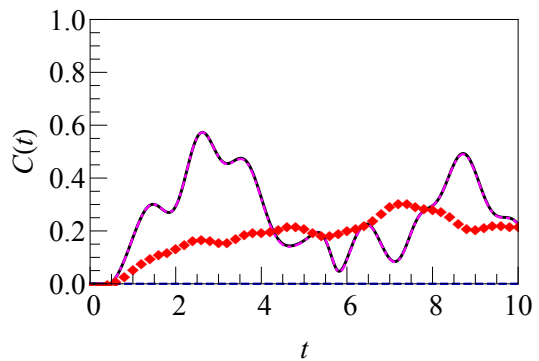}
\caption{(Color online) Dynamics of entanglement between spins 1 and 4 for the XYZ model with $N = 4$. We have used $\lambda_1 = 0.5, \lambda_2 = 1, \lambda_3 = 0.25$. We have shown the dynamics without any control fields (dashed, blue line), with control fields using the total Hamiltonian $H_{0}+H_{\text{SB}}+H_{c}(t)$ (dot-dashed, magenta curve), and using the time-averaged Hamitonians $\bar{H}_1$ (red diamonds) and $\bar{H}_2$ (solid, black curve). The initial state of the spin chain is $\ket{0000}$. The dashed, blue curve again overlaps with the horizontal axis.}
\label{Hbar1Hbar2XYZfigure}
\end{figure}

We now look at the full XYZ spin chain Hamiltonian, given by
\begin{dmath*}
H_{0}=\sum_{k=1}^{3} \lambda_{k}\sum_{j=1}^{N-1}\sigma_{k}^{(j)}\sigma_{k}^{(j+1)}.
\end{dmath*} 
One important comment is in order. If the spin chain is fully isotropic, that is, $\lambda_1 = \lambda_2 = \lambda_3$, then the control fields cannot alter the spin chain Hamiltonian. The reason is simple - for the fully isotropic case, the Hamiltonian can be written as an inner product of a vector consisting of the Pauli matrices with itself, and this inner product is of course invariant under unitary operations. Therefore, we will focus on the anisotropic case where the coupling strengths are not all equal to each other. The time-averaged Hamiltonian is given by $\bar{H}_1$ if $n_y \neq 2n_x$ [see Eq.~\eqref{Hbar1eq}] and by $\bar{H}_2$ [see Eq.~\eqref{Hbar2eq}] if $n_y = 2n_x$. As shown in Fig.~\ref{Hbar1Hbar2XYZfigure}, with the initial state $\ket{0000}$, the dynamics are captured very well by our effective Hamiltonian since the dot-dashed magenta and solid, black curves overlap. It is also clear that the control fields with $n_y = 2n_x$ are better at generating entanglement for the given values of the interaction strengths. Once again, the choice of the control fields can affect the interactions, and thereby the generation of a quantum resource such as entanglement, to a very large degree. Similar conclusions hold true if we consider the initial state to be $\ket{0111}$, $\ket{1000}$, or $\ket{1111}$ instead.

\section{Comparing the effective Hamiltonians $\bar{H}_1$ and $\bar{H}_2$ for larger $N$}
Having demonstrated that if we apply sufficiently strong and rapidly oscillating control fields, the spin chain Hamiltonian can be approximated by $\bar{H}_1$ if $n_y \neq 2n_x$ and by $\bar{H}_2$ if $n_y = 2n_x$, we now aim to cast $\bar{H}_1$ and $\bar{H}_2$ in diagonal form so that the concurrence for larger spin chains can be worked out easily. A commonly used tool in such a calculation is the Jordan-Wigner transformation which allows one to transform the problem of interacting spins to a problem of spinless fermions \cite{LiebMattisarticle, McCoyXY1}. Unfortunately, the presence of the $\sigma_z^{(j)} \sigma_z^{(j+1)}$ interactions in $\bar{H}_1$ and $\bar{H}_2$ means that the spinless fermions are interacting. However, if the original spin chain is such that $2\lambda_1 + \lambda_2 + \lambda_3 = 0$, then the Jordan-Wigner transformation allows us to transform both $\bar{H}_1$ and $\bar{H}_2$ to non-interacting fermions, thereby making the problem tractable and greatly reducing the computational complexity. We largely follow the approach presented in Refs.~\cite{LiebMattisarticle,AmicoJPA2004}, although we must emphasize that the spin chain Hamiltonian $\bar{H}_2$ we are solving is different.

\begin{figure}[h!]
\centering
\includegraphics[width=.8\linewidth]{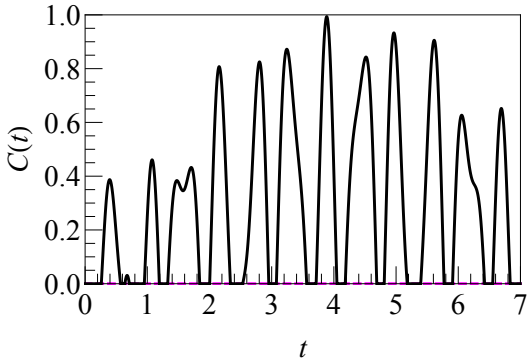}
\caption{(Color online) Dynamics of entanglement between spins 1 and 4 for the XYZ model with $N = 4$. We have used $\lambda_1=2, \lambda_2=1, \lambda_3 =-5$. The solid, black curve shows the entanglement with $\bar{H}_2$, while the dashed magenta curve is for $\bar{H}_1$. The dashed magenta curve overlaps with the horizontal axis. The initial state of the spin chain is $\ket{1111}$. We have checked that when the concurrence is approximately one, the purity of the state for spins $1$ and $4$ is very close to one with the two-spin state being approximately the entangled state $0.75 \ket{00} + (-0.55 + 0.37i)\ket{11}$.}
\label{Hbar1Hbar2XYZspecial}
\end{figure}

\begin{figure}[h!]
\centering
\includegraphics[width=.8\linewidth]{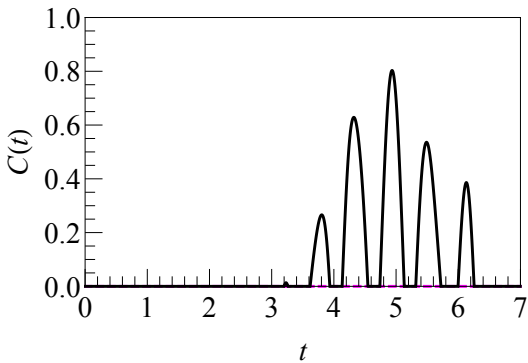}
\caption{(Color online) Dynamics of the entanglement between spins 1 and 12 for the XYZ model with $N = 12$. We have used $\lambda_{1}=2,\lambda_{2}=1$, and $\lambda_3 = -5$. The solid black curve shows the concurrence with the time-averaged Hamiltonian $\bar{H}_2$, while the dashed magenta curve shows the dynamics with $\bar{H}_1$. The dashed magenta curve again overlaps with the horizontal axis. The initial state of the spin chain is $\ket{11\hdots 1}$.}
\label{Hbarlargespinchain1and12}
\end{figure}

\begin{figure}[h!]
\centering
\includegraphics[width=.8\linewidth]{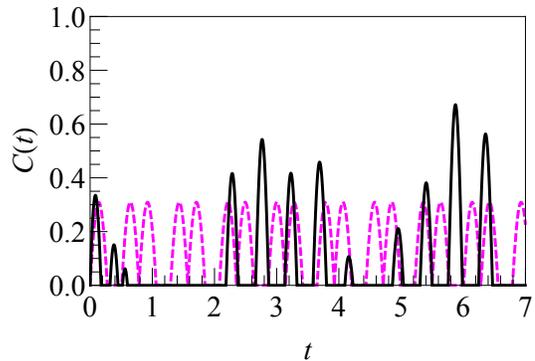}
\caption{(Color online) Same as Fig.~\ref{Hbarlargespinchain1and12}, except that we are now looking at the concurrence between spins 6 and 7.}
\label{Hbarlargespinchain6and7}
\end{figure}

With the condition $\lambda_3 = -2\lambda_1 - \lambda_2$ to suppress the interaction terms $\sigma_{z}^{(j)}\sigma_{z}^{(j+1)}$, we find that the effective Hamiltonians are
\begin{dmath*}
\bar{H}_1=-\lambda_{1}\sum_{j=1}^{N-1}\sigma_{y}^{(j)}\sigma_{y}^{(j+1)},
\end{dmath*}
while 
\begin{dmath*}
\bar{H}_2 = \sum_{j=1}^{N-1}\left\lbrace\frac{(\lambda_{1}+\lambda_{2})}{2}\left[\sigma_{x}^{(j)}\sigma_{y}^{(j+1)}+\sigma_{y}^{(j)}\sigma_{x}^{(j+1)}\right]-\lambda_{1}\left[\sigma_{y}^{(j)}\sigma_{y}^{(j+1)}\right]\right\rbrace.
\end{dmath*}
The details of finding the dynamics with these effective Hamiltonians are given in Appendix \ref{Jordanwignerappendix}. In summary, to find the entanglement between any two spins of the spin chain, we first need to find the two-spin density matrix. In order to find the two-spin density matrix, we calculate the correlation functions. These correlation functions can be expressed in terms of Pfaffians as discussed in Appendix \ref{Jordanwignerappendix}. Using this approach, we have checked that for small spin chains, the results produced are the same as those obtained numerically. For example, the results shown in Fig.~\ref{Hbar1Hbar2XYZspecial} were reproduced using the approach employing the Jordan-Wigner transformation. Interestingly, in this case, with the initial state $\ket{1111}$, we can achieve almost perfect entanglement between the first and last spins with $\bar{H}_2$, while $\bar{H}_1$ generates no entanglement at all. In Appendix \ref{appendixny2nx}, we investigate how the entanglement generated changes as the difference between $n_y$ and $2n_x$ changes. We then used the Jordan-Wigner transformation approach to obtain the concurrence for larger spin chains. As an example, in Fig.~\ref{Hbarlargespinchain1and12} we have shown the dynamics of the entanglement between the ends of a spin chain with $N = 12$ for both $\bar{H}_1$ and $\bar{H}_2$ with the initial state $\ket{11\hdots 1}$. It is clear that the entanglement generated if the special control fields with $n_y = 2n_x$ are used is considerable, while no entanglement is generated when $n_y \neq 2n_x$. As the size of the spin chain is increased further, we found that, with $\bar{H}_2$, the entanglement between the ends of the spin chain decreases and the time at which this maximum is obtained increases, while no entanglement is generated with $\bar{H}_1$. For example, with $N = 24$, the maximum concurrence obtained between the ends with $\bar{H}_2$ is approximately $0.65$ for $t = 8.7$. It should also be kept in mind that with large spin chains, as the number of spins is increased, we should consider a rescaled concurrence that takes into account the number of spins. For example, in Ref.~\cite{VidalPRA2004b}, the rescaled concurrence has been defined as $C_R = (N - 1)C$. It is then clear that the entanglement generated by the spin chain dynamics is very significant if control fields with $n_y = 2n_x$ are used. Moreover, if we look instead at, for instance, spins 6 and 7 of the spin chain, we again observe that $\bar{H}_2$ is better at generating entanglement as compared to $\bar{H}_1$ (see Fig.~\ref{Hbarlargespinchain6and7}).

\section{Conclusion}

In summary, we have shown that applying suitable control fields to a spin chain can, at least to a large extent, eliminate the interaction between the spin chain and its environment. Moreover, we have also shown that the application of the control fields modulates the spin-spin interaction in ways that can effectively generate spin-spin interactions that are absent in the original spin chain Hamiltonian. As an example of the constructive use of this modification, we have shown how, starting from the quantum Ising chain, perfect quantum state transfer can be achieved provided that control fields of sufficient strength and frequency are applied. We have also presented numerical simulations which show that two-spin entanglement generation in the spin chain can be improved by using particular control fields. Moreover, we have also diagonalized the effective spin chain Hamiltonian, at least for special coupling strengths, to show how the effect of the control fields can be analyzed for larger spin chains. Due to the great theoretical and experimental interest in spin chains, especially in the context of quantum computation and information, our results should be interesting not only from the perspective of effectively isolating spin chains from their environment, but also for engineering spin-spin interactions via simple static and oscillating control fields.   

\section*{acknowledgements}
The authors acknowledge support from the LUMS FIF Grant FIF-413. A.~Z.~C. is also grateful for support from HEC under grant No 5917/Punjab/NRPU/R\&D/HEC/2016. Support from the National Center for Nanoscience and Nanotechnology is also acknowledged.

\appendix

\section{Eliminating the spin chain-environment interaction and transformation of the spin chain Hamiltonian}
\label{eliminatinginteraction}

\begin{figure}[h!]
\centering
\includegraphics[width=.8\linewidth]{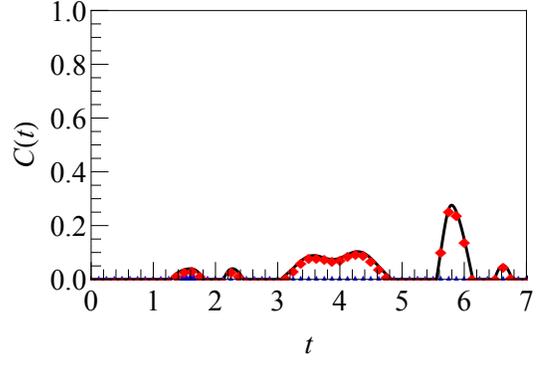}
\caption{(Color online) Dynamics of entanglement between spins 1 and 4 for the XY model with $N = 4$ in the presence of control fields. We have used $\lambda_1 = 1, \lambda_2 = 1$, and $\lambda_3 = 0$. The solid, black curve shows the entanglement with the effective Hamiltonian $\bar{H}_2$, while the red diamonds show the concurrence with the full time-dependent Hamiltonian [$H_0 + H_c(t) + H_{\text{SB}}(t)$] with $t_c = 0.01$. The blue triangles shows the dynamics of the full time-dependent Hamiltonian with $t_c = 0.5$. Here $n_y = 2$ and $n_x = 1$. The solid, black curve and the red diamonds overlap, while the blue triangles largely overlap with the horizontal axis. As always, the effect of the environment is modeled by Ornstein-Uhlenbeck processes with zero mean, correlation time $\tau = 0.5$, and standard deviation $\sigma = 2.0$. The initial state used here is $\ket{1000}$.}
\label{labeltctoolarge}
\end{figure}

Let us write the Hamiltonian for the spin chain in the presence of the control fields as 
\begin{dmath*}
H_{\text{tot}}=H_{0}+H_{c}(t)+H_{B}+H_{\text{SB}}=H^{\prime }+ H_B + H_{c}(t).
\end{dmath*}
Here $H_c(t)$ is the control field Hamiltonian (acting on the spin chain), $H_B$ is the Hamiltonian of the environment, and $H_{\text{SB}}$ is the interaction between the spin chain and its environment. It is interesting to note that the environment of a quantum system itself has been modeled as a spin chain (see, for instance, Refs.~\cite{cucchietti2005decoherence,JafariPRB2017,ChaudhryEurJPhysD2019} and references therein). For future convenience, we have defined
\begin{dmath}
H'=   H_{0} + H_{\text{SB}}.
\end{dmath}
Our goal is to see how a state evolves under the action of the
total Hamiltonian. To this end, let us first rotate the basis by $U_{c}(t)U_B(t)$, where $U_B(t) = e^{-iH_Bt}$. In this frame, the unitary time-evolution operator for the spin chain and the environment as a whole is
\begin{dmath*}
\tilde{U}_{\text{tot}}(t) =\mathcal{T}\text{exp}\left[-i \int_{0}^{t}      \tilde{H}^{\prime }(s)ds \right],
\end{dmath*}
where $\tilde{H}^{\prime }(s)=U_B^\dagger (s) U_{c}^{ \dagger}(s)H^{\prime }U_{c}(s)U_B(s)$. At time $t = Nt_{c}$ ($N$ is a positive integer), due to the periodicity of the control fields,
\begin{dmath*}
\tilde{U}_{\text{tot}}(t)=\left[\tilde{U}_{\text{tot}}(t_{c})\right]^{N},
\end{dmath*}
and
\begin{dmath*}
\tilde{U}_{\text{tot}}(t_{c}) =\mathcal{T}\text{exp}\left[-i \int_{0}^{t_{c}}      \tilde{H}^{\prime }(s) \,ds \right].
\end{dmath*}
Now comes the key step. We use the Magnus expansion to write
 \begin{dmath*}
\tilde{U}_{\text{tot}}(t_{c})=\text{exp} [ -it_{c}(\tilde{H}^{(0)} +\tilde{H}^{(1)} + \cdots) ],
\end{dmath*}
where 
$$ \tilde{H}^{(0)} = \frac{1}{t_c}\int_0^{t_c} \tilde{H}'(t) \, dt, $$
and 
$$ \tilde{H}^{(1)} = -\frac{i}{2t_c} \int_0^{t_c} dt_1 \int_0^{t_1} dt_2 \, [\tilde{H}'(t_1), \tilde{H}'(t_2)]. $$
$\tilde{H}^{(0)}$ is independent of $t_c$, while $\tilde{H}^{(1)}$ increases with increasing $t_c$, suggesting that for small $t_c$, only $\tilde{H}^{(0)}$ can be considered. Now, 
\begin{dmath*}
\tilde{H}^{(0)} = \frac{1}{t_c}\int_0^{t_c} U_c^\dagger(t) H_0 U_c(t) \, dt + \frac{1}{t_c} \int_0^{t_c} U_c^\dagger (t) U_B^\dagger(t) H_{\text{SB}} U_B(t) U_c(t) \, dt. 
\end{dmath*}
Writing $H_{\text{SB}} = \sum_\alpha F_\alpha \otimes B_\alpha$, the latter term can be written as 
$$ \frac{1}{t_c} \int_0^{t_c} \sum_\alpha U_c^\dagger (t) F_\alpha U_c(t) U_B^\dagger(t) B_\alpha U_B(t) \, dt. $$
If $t_c$ is much smaller than the environment correlation time $\tau$, that is $\frac{t_c}{\tau} \ll 1$, then the time dependence of the environment operators over the timescale $t_c$ can be neglected, leading to 
\begin{dmath*}
\int_0^{t_c} \sum_\alpha U_c^\dagger (t) F_\alpha U_c(t) B_\alpha  \, dt = \int_{0}^{t_{c}}U_{c}^{ \dagger}(t)H_{\text{SB}}U_{c}(t) dt.
\end{dmath*}
If we now impose the condition 
\begin{dmath*}
\int_{0}^{t_{c}}U_{c}^{ \dagger}(t)H_{\text{SB}}U_{c}(t) dt=0,
\end{dmath*}
we find that 
\begin{dmath*}
\tilde{H}^{(0)} = \frac{1}{t_c}\int_0^{t_c} U_c^\dagger(t) H_0 U_c(t) \, dt.
\end{dmath*}
Considering only the first term in the Magnus expansion, 
\begin{equation*}
\tilde{U}_{\text{tot}}(t) \approx [e^{-it_{c}\tilde{H}^{(0)}}]^{t/t_{c}}=e^{-it \bar{H}},
\end{equation*}
with 
$$ \bar{H} = \frac{1}{t_c} \int_0^{t_c} U_c^\dagger(t) H_0 U_c(t) \, dt. $$ 
Transforming it back to the original frame, we find that the unitary evolution operator is
\begin{dmath}
U_{\text{tot}}(t) \approx U_{c}(t) U_B(t) e^{-it \bar{H}}.
\end{dmath}
Thus, the spin chain and its environment have been effectively decoupled, at least to lowest order. Note that $t_c$ has to be much smaller than the environment correlation time $\tau$ for our scheme to work (a similar result holds when pulses are used - see, for instance, Ref.~\cite{KurizkiPRA20112}). As an illustration, in Fig.~\ref{labeltctoolarge} we have shown the concurrence obtained with $t_c = 0.01$ and $t_c = 0.5$, and the environment correlation time is $\tau = 0.5$. It is clear that with $t_c = 0.5$, the control fields cannot protect the spin chain against the effect of the environment.

It is also worth examining the next term in the Magnus expansion. The part of $\tilde{H}^{(1)}$ concerned with the dynamics of the spin chain only is 
$$ -\frac{i}{2t_c} \int_0^{t_c} dt_1 \int_0^{t_1} dt_2 \, [\tilde{H}_0(t_1), \tilde{H}_0(t_2)]. $$
This is proportional to $\lambda_{jk}^2 t_c$, while $\bar{H}$ is proportional to $\lambda_{jk}$. Thus, the correction to the effective Hamiltonian is negligible if $\lambda_{jk} t_c \ll 1$, with our results becoming exact in the limit $\lambda_{jk} t_c \rightarrow 0$. To see this more concretely, consider $H_0 = \lambda \sigma_x^{(1)} \sigma_x^{(2)}$. Now,
\begin{align*}
&U_c^\dagger (t) \sigma_x^{(1)} \sigma_x^{(2)} U_c(t) = \cos^2(2\omega n_y t) \sigma_x^{(1)}\sigma_x^{(2)} + \\ 
&\frac{1}{2} \sin(4\omega n_y t) \sigma_x^{(1)} \sigma_y^{(2)} + \frac{1}{2} \sin(4\omega n_y t) \sigma_y^{(1)} \sigma_x^{(2)} + \\
&\sin^2(2\omega n_y t) \sigma_y^{(1)} \sigma_y^{(2)}. 
\end{align*}
We can then work out $ -\frac{i}{2t_c} \int_0^{t_c} dt_1 \int_0^{t_1} dt_2 \, [\tilde{H}_0(t_1), \tilde{H}_0(t_2)]$. Although the full expression is long and cumbersome, it is clear that one of the terms is 
\begin{align*}
-\frac{i}{2t_c} \int_0^{t_c} dt_1 \int_0^{t_1} dt_2 \frac{\lambda^2}{2} \cos^2 (2\omega n_y t_1) \sin(4\omega n_y t_2) \times \\
[\sigma_x^{(1)} \sigma_x^{(2)}, \sigma_x^{(1)}\sigma_y^{(2)}]. 
\end{align*}
For $n_y = 2$, this is equal to $\frac{\lambda^2 t_c}{128\pi}\sigma_z^{(2)}$, thus illustrating our claim that $\tilde{H}^{(1)}$ contains terms proportional to $t_c$. In short, the spin chain is effectively decoupled from the environment and its dynamics can be obtained from the effective Hamiltonian if $\frac{t_c}{\tau} \ll 1$ and $\lambda_{j,k} t_c \ll 1$. The latter condition is similar to what has been obtained before when pulses are applied to the spin chain \cite{ChoiPRL2017}. In all the  simulations demonstrating the usefulness of our control fields, these conditions are satisfied.

\section{Including effective magnetic fields in the effective Hamiltonian}
\label{includingstaticfields}

Consider the total Hamiltonian 
\begin{equation}
H = H_0 + H_c(t) + H_{\text{SB}} + H_d(t),
\label{totalHwithaddfields}
\end{equation}
where $H_0$, $H_{\text{SB}}$, and $H_c(t)$ given by Eqs.~\eqref{originalspinchainHamiltonian}, \eqref{systemenvironmentinteraction}, and \eqref{controlHamiltonian} respectively. $H_d(t)$ is an additional Hamiltonian of the form
\begin{dmath}
\label{additionalfieldsforstatic}
$$H_d(t) = \sum_{j = 1}^N \left[ b_{j1} \cos(2\omega n_y t) \sigma_x^{(j)} + b_{j2} \cos(2\omega n_x t) \sigma_y^{(j)} + b_{j3} \sin(2\omega n_z t) \sigma_z^{(j)}\right], $$
\end{dmath}
where $n_x$ and $n_y$ are the same integers as in $H_c(t)$, while $n_z$ is also an integer. Once again transforming to the frame of the control fields, we find that the effective Hamiltonian is now 
$$\bar{H} = \frac{1}{t_c} \int_0^{t_c} \, dt \, U_c^\dagger(t) H_0 U_c(t) + \frac{1}{t_c} \int_0^{t_c} \, dt \, U_c^\dagger(t) H_d(t) U_c(t). $$
The first term leads to the same effective Hamiltonian as before. Thus, an additional term has been added to the effective time-independent Hamiltonian. Denoting $\bar{H}_d = \frac{1}{t_c} \int_0^{t_c} \, dt \, U_c^\dagger(t) H_d(t) U_c(t)$, we find in a straightforward manner that 
\begin{dmath}
\label{effectivewithstatic1}
\bar{H}_d = \sum_{j = 1}^N \left[ \frac{b_{j1}}{2} \sigma_x^{(j)} + \frac{b_{j2}}{2} \sigma_y^{(j)} + \frac{b_{j3}}{4}\sigma_z^{(j)} \right], 
\end{dmath}
for $n_y \neq 2n_x$ and $n_z = n_y - n_x$. If, on the other hand, $n_y = 2n_x$ and $n_z = n_y - n_x$, we find that 
\begin{dmath}
\label{effectivewithstatic2}
\bar{H}_d = \sum_{j = 1}^N \left[ \frac{b_{j1}}{2} \sigma_x^{(j)} +  \frac{b_{j2}}{4} \sigma_x^{(j)} + \frac{b_{j2}}{2} \sigma_y^{(j)} + \frac{b_{j3}}{4}\sigma_z^{(j)} \right].
\end{dmath}
In this way, by applying additional oscillating fields with frequencies similar to the control fields in $H_c(t)$, we can also effectively add static magnetic fields to the spin chain Hamiltonian. Similar to our prior treatment, we expect our effective Hamiltonian approach to be valid if $b_{jk} t_c \ll 1$. As an example, if we choose $b_{j1} = b_{j3} = 0$, and we start from the quantum Ising model $H_0 = \sum_{j = 1}^{N -1} \lambda_j \sigma_x^{(j)} \sigma_x^{(j+1)}$, by applying the control fields as well as additional oscillating fields described by $H_d(t) = \sum_{j = 1}^N b_{j} \cos(2\omega n_x t) \sigma_y^{(j)}$, we end up with effective Hamiltonian (taking $n_y \neq 2n_x$)
\begin{dmath}
\bar{H} = \sum_{j = 1}^{N-1} \frac{\lambda_j}{2} \left[\sigma_x^{(j)} \sigma_x^{(j+1)} + \sigma_z^{(j)} \sigma_z^{(j+1)}\right] + \sum_{j=1}^N \frac{b_j}{2} \sigma_y^{(j)}.
\end{dmath}
This is effectively the quantum XX model with a transverse magnetic field. As another example, we can start from the anisotropic Heisenberg Hamiltonian $H_0 = \sum_{j = 1}^{N - 1} \left[\lambda_1 \sigma_x^{(j)} \sigma_x^{(j+1)} + \lambda_2 \sigma_y^{(j)} \sigma_y^{(j+1)} + \lambda_3 \sigma_z^{(j)} \sigma_z^{(j+1)} \right]$. With the application of the control Hamiltonian as well as the additional oscillating field described by $H_d = \sum_{j = 1}^N b \sin(2\omega n_z t) \sigma_z^{(j)}$, the effective Hamiltonian is (with $n_y = 2n_x$ and $n_z = n_y - n_x$) 
\begin{dmath}
\bar{H} = \sum_{j=1}^{N-1}\left\lbrace \frac{\lambda_{1}}{2}\left[\sigma_{x}^{(j)}\sigma_{x}^{(j+1)}+\sigma_{z}^{(j)}\sigma_{z}^{(j+1)}\right]+ \frac{\lambda_{j}}{4}\left[\sigma_{x}^{(j)}\sigma_{x}^{(j+1)}+2\sigma_{y}^{(j)}\sigma_{y}^{(j+1)}+\sigma_{x}^{(j)}\sigma_{y}^{(j+1)}+\sigma_{y}^{(j)}\sigma_{x}^{(j+1)}+\sigma_{z}^{(j)}\sigma_{z}^{(j+1)}\right]+\frac{\lambda_{3}}{4}\left[\sigma_{x}^{(j)}\sigma_{x}^{(j+1)}+2\sigma_{y}^{(j)}\sigma_{y}^{(j+1)}-\sigma_{x}^{(j)}\sigma_{y}^{(j+1)}-\sigma_{y}^{(j)}\sigma_{x}^{(j+1)}+\sigma_{z}^{(j)}\sigma_{z}^{(j+1)}\right]\right\rbrace + \sum_{j = 1}^N \frac{b}{4}\sigma_z^{(j)}.
\label{effectivestaticfieldHamiltonian}
\end{dmath}
To illustrate our results, in Fig.~\ref{labeladdstaticfield} below, we have plotted the concurrence between the first and last spins for $N = 4$ using the effective Hamiltonian Eq.~\eqref{effectivestaticfieldHamiltonian} and using the total Hamiltonian Eq.~\eqref{totalHwithaddfields} with $b_{j1} = b_{j2} = 0$, while $b_{j3} = b$ for all $j$. It is clear that the effective Hamiltonian captures the exact dynamics exceedingly well.

\begin{figure}[h!]
\centering
\includegraphics[width=.8\linewidth]{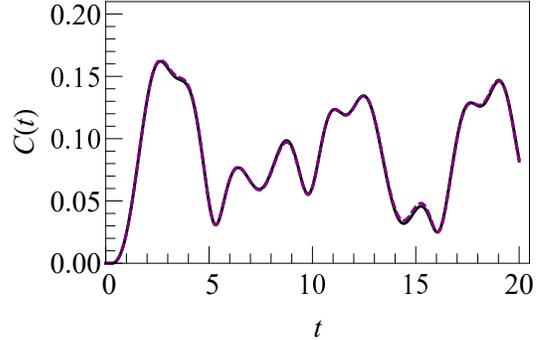}
\caption{(Color online) Dynamics of entanglement between spins 1 and 4 for the XYZ model with $N = 4$ in the presence of an additional oscillating field. We have used $\lambda_1 = 0.2, \lambda_2 = 0.4, \lambda_3 = 0.3$, and $b_{j1} = b_{j2} = 0$, while $b_{j3} = 1$ for all $j$. The solid, black curve shows the entanglement with the effective Hamiltonian $\bar{H}$ given by Eq.~\eqref{effectivestaticfieldHamiltonian}, while the dashed, purple curve shows the concurrence with the full Hamiltonian given in Eq.~\eqref{totalHwithaddfields}. Here $n_y = 2$ and $n_x = 1$. The curves overlap. The initial state is $\ket{0000}$.}
\label{labeladdstaticfield}
\end{figure}

\section{Simulating the effect of noise}
\label{appendixnoise}

In the numerical simulations, the effect of the environment on the spin chain is modeled by the Hamiltonian 
$$H_{\text{SB}} = \sum_{j = 1}^N B_x \sigma_x^{(j)} + B_y \sigma_y^{(j)} + B_z \sigma_z^{(j)}, $$
where $B_x$, $B_y$, and $B_z$ are independent random variables obtained by solving the Ornstein-Uhlenbeck equation \cite{Jacobsbook} cast in the form  
$$ dB_m = -\frac{(B_m - \mu)}{\tau} dt + \sigma \sqrt{\frac{2}{\tau}} dW. $$
Here $m = x, y, z$, $\mu$ is the mean, $\sigma$ is the standard deviation, and $\tau$ is the correlation time. $W$ is the standard Wiener process. Note that $\mu$ and $\sigma$ have the same dimensions as $B_m$, and since this is a stochastic differential equation, $(dW)^2 = dt$. Throughout the paper, we have used $\sigma = 2.0$ (which is comparable to the spin-spin coupling strengths; see Fig.~\ref{statetransferN4} caption), $\mu = 0$ (that is, the mean is negligible compared to the spin-spin coupling strengths), and $\tau = 0.5$ (since we are using $t_c = 0.01$ throughout, this means that $\frac{t_c}{\tau} = 0.02$, thus fulfilling our dynamical decoupling condition). It should be kept in mind that in our system of units with $\hbar = 1$, $B_m$ (and thus $\sigma$) and the spin-spin coupling strengths have the same units [see Eqs.~\eqref{originalspinchainHamiltonian} and \eqref{systemenvironmentinteraction}].

\section{Implementing single-spin operations}
\label{appendixsinglespingate}

\begin{figure}[t!]
\centering
\includegraphics[width=.8\linewidth]{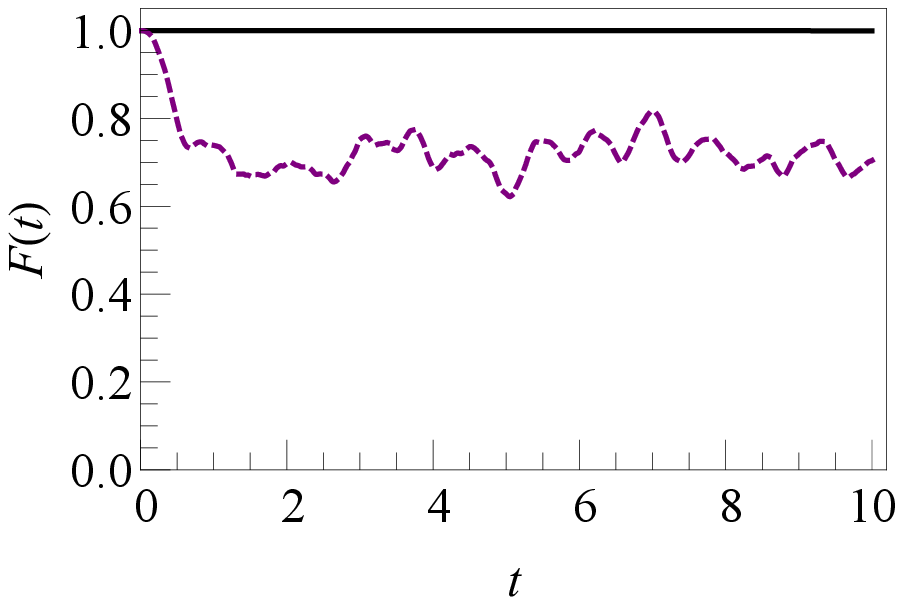}
\caption{(Color online) Dynamics of the fidelity of the first spin with $N = 4$. We have used $\lambda_1 = 0.2, \lambda_2 = 0.4$, and $\lambda_3 = 0.5$. The solid, black curve shows the fidelity in the presence of the control fields given by Eq.~\eqref{modifiedcontrolHamiltonian} ($n_x^{(1)} = 4$, $n_y^{(1)} = 8$, $n_x^{(2)} = 1$, and $n_y^{(2)} = 2$ with $t_c = 0.01$), while the dashed, purple curve shows the fidelity for the first spin in the absence of any control fields. The initial state of the spin chain is $\ket{0000}$. As before, the effect of the environment is modeled by Ornstein-Uhlenbeck processes with $\sigma = 2.0$ and $\tau = 0.5$.}
\label{labelspinspindecoupling}
\end{figure}

The dynamical decoupling approach can be adapted to implement single qubit operations (see, for example. Refs.~\cite{ViolaPRL2009, ChaudhryPRA12012}), which we now do in the context of spin chains for continuous control fields. Without loss of generality, let us suppose that we require a desired unitary operation to be implemented on the first spin in the spin chain (similar considerations apply if the unitary operation is to be applied on some other spin in the spin chain). To implement such a single-spin operation, we first find continuous fields that not only remove the effect of $H_{\text{SB}}$, but also remove the effect of the spin-spin coupling between the first spin and its nearest neighbor. That is, we need to find $U_c(t)$ such that not only $\int_0^{t_c} \, dt\, U_c^\dagger (t) H_{\text{SB}} U_c(t) = 0$, but also that $\int_0^{t_c} \, dt \, U_c^\dagger (t) \sigma_x^{(1)} \sigma_x^{(2)} U_c(t) = \int_0^{t_c} \, dt \, U_c^\dagger (t) \sigma_y^{(1)} \sigma_y^{(2)} U_c(t) = \int_0^{t_c} \, dt \, U_c^\dagger (t) \sigma_z^{(1)} \sigma_z^{(2)} U_c(t) = 0$. To achieve this, we modify $U_c(t)$ to 
\begin{equation}
\label{modifiedUc}
U_c(t) = e^{i\omega n_x^{(1)}\sigma_x^{(1)}t} e^{i\omega n_y^{(1)}\sigma_y^{(1)}t} \prod_{i = 2}^N e^{i\omega n_x^{(2)}\sigma_x^{(i)}t} e^{i\omega n_y^{(2)}\sigma_y^{(2)}t}. 
\end{equation}
This means that we apply different fields to the first spin as compared to all the other spins, that is,
\begin{dmath}
\label{modifiedcontrolHamiltonian}
H_{c}(t)=\sum_{i=1}^{N}\left\lbrace \omega n_y^{(i)} [\sin(2 \omega n_x^{(i)} t )\sigma_{z}^{(i)} - \cos(2  \omega n_x^{(i)} t)\sigma_{y}^{(i)}] -\omega n_x^{(i)} \sigma_{x}^{(i)} \right\rbrace,
\end{dmath}
with $n_x^{(2)} = n_x^{(3)} = \hdots = n_x^{(N)}$, and $n_y^{(2)} = n_y^{(3)} = \hdots = n_y^{(N)}$. From the condition $\int_0^{t_c} \, dt\, U_c^\dagger (t) H_{\text{SB}} U_c(t) = 0$, we find that $n_x^{(1)} \neq n_y^{(1)}$ and $n_x^{(2)} \neq n_y^{(2)}$. Next, $\int_0^{t_c} \, dt \, U_c^\dagger (t) \sigma_x^{(1)} \sigma_x^{(2)} U_c(t) = 0$ if $n_y^{(2)} \neq n_y^{(1)}$. The requirement that $\int_0^{t_c} \, dt \, U_c^\dagger (t) \sigma_y^{(1)} \sigma_y^{(2)} U_c(t) = 0$ leads to the following conditions on the control fields:
\begin{align*}
n_x^{(1)} &\neq n_x^{(2)}, \\
n_x^{(1)} + n_x^{(2)} &\neq n_y^{(2)}, \\
n_x^{(1)} - n_x^{(2)} &\neq n_y^{(2)}, \\
n_x^{(2)} - n_x^{(1)} &\neq n_y^{(2)}, \\
n_x^{(1)} + n_x^{(2)} &\neq n_y^{(1)}, \\
n_x^{(1)} - n_x^{(2)} &\neq n_y^{(1)}, \\
n_x^{(2)} - n_x^{(1)} &\neq n_y^{(1)}, \\
n_x^{(1)} - n_x^{(2)} + n_y^{(1)} + n_y^{(2)} &\neq 0, \\
n_x^{(1)} - n_x^{(2)} - n_y^{(1)} - n_y^{(2)} &\neq 0,\\
n_x^{(1)} - n_x^{(2)} + n_y^{(1)} - n_y^{(2)} &\neq 0,\\
n_x^{(1)} - n_x^{(2)} - n_y^{(1)} + n_y^{(2)} &\neq 0,\\
n_x^{(1)} + n_x^{(2)} - n_y^{(1)} - n_y^{(2)} &\neq 0,\\
n_x^{(1)} + n_x^{(2)} + n_y^{(1)} - n_y^{(2)} &\neq 0,\\
n_x^{(1)} + n_x^{(2)} - n_y^{(1)} + n_y^{(2)} &\neq 0.
\end{align*}
Setting $\int_0^{t_c} \, dt \, U_c^\dagger (t) \sigma_z^{(1)} \sigma_z^{(2)} U_c(t) = 0$ leads to the same conditions. It is then straightforward to find integers $n_x^{(1)}$, $n_y^{(1)}$, $n_x^{(2)}$, and $n_y^{(2)}$ that satisfy these requirements. For example, $n_x^{(1)} = 4$, $n_y^{(1)} = 8$, $n_x^{(2)} = 1$, and $n_y^{(2)} = 2$ do the job. We illustrate the performance of these control fields in protecting the state of the first spin in Fig.~\ref{labelspinspindecoupling}. It is clear that our control fields are able to preserve the state of the first spin. 

\begin{figure}[b!]
\centering
\includegraphics[width=.8\linewidth]{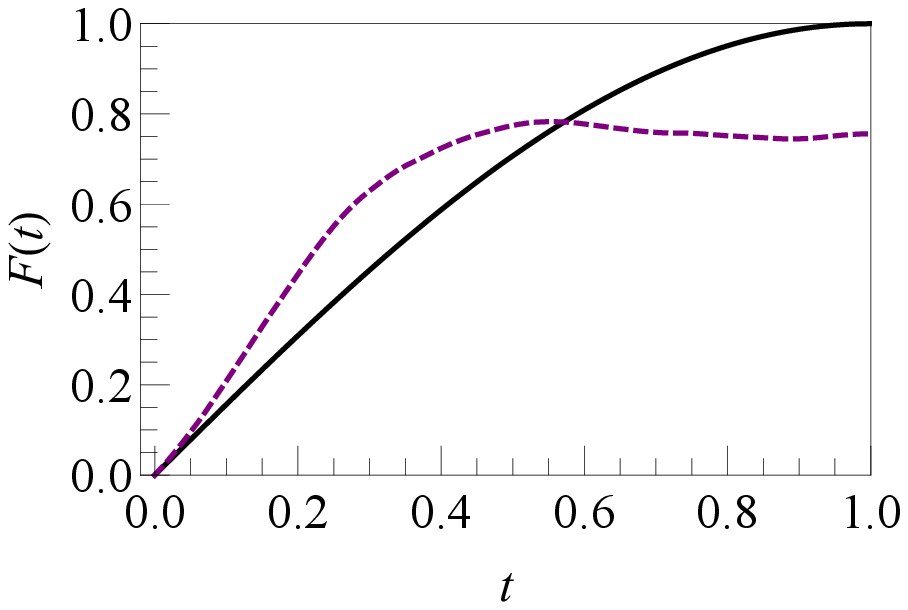}
\caption{(Color online) Dynamics of the fidelity of the first spin with $N = 4$. We have used $\lambda_1 = 0.2, \lambda_2 = 0.4$, and $\lambda_3 = 0.5$. The solid, black curve shows the fidelity in the presence of the control fields given by Eq.~\eqref{Hgate} with $n_x^{(1)} = 4$, $n_y^{(1)} = 8$, $n_x^{(2)} = 1$, $n_y^{(2)} = 2$, and $t_c = 0.01$. Here we have used $t_g = 1$. The dashed, purple curve shows the fidelity for the first spin if the single spin operation is implemented via the simple control field $-\frac{\pi}{2t_g}\sigma_x^{(1)}$ in the presence of noise. The initial state of the spin chain is $\ket{0000}$.}
\label{labelsinglequbitgate}
\end{figure} 

We now implement the single spin operation on the first spin. As an example, consider the unitary operation $e^{i\frac{\pi}{2t_g}\sigma_x^{(1)} t}$. This transforms the state $\ket{0}$ for the first spin to the state $\ket{1}$ after time $t_g$. To implement this operation, we first transform to the frame of the control fields. Then, we implement the single spin operation in this frame, and thereafter transform back to the original frame. The net result is that the unitary operator that needs to be implemented is given by 
$$ U_{\text{gate}}(t) = U_c(t) e^{i\frac{\pi}{2t_g}\sigma_x^{(1)} t}, $$
with $U_c(t)$ given by Eq.~\eqref{modifiedUc}.
The corresponding Hamiltonian $H_{\text{gate}}(t)$ is obtained via $H_{\text{gate}}(t) = i\frac{\partial U_{\text{gate}}}{\partial t} U_{\text{gate}}^\dagger$. A simple calculation shows that 
\begin{dmath}
\label{Hgate}
H_{\text{gate}}(t) = H_c(t) - \frac{\pi}{2t_g}\left[\sigma_x^{(1)} \cos(2\omega n_y^{(1)} t) + \sigma_z^{(1)} \sin(2\omega n_y^{(1)}t) \cos(2\omega n_x^{(1)} t) + \sigma_y^{(1)} \sin(2\omega n_y^{(1)} t) \sin(2\omega n_x^{(1)} t)\right],
\end{dmath} 
with $H_c(t)$ given by Eq.~\eqref{modifiedcontrolHamiltonian}. With these albeit complicated control fields, we are able to implement, at least in principle, a single spin operation with high fidelity. The performance of such a single spin protected gate is illustrated in Fig.~\ref{labelsinglequbitgate}, from which it is clear that high fidelities can indeed be achieved.

\section{The Jordan-Wigner Transformation}
\label{Jordanwignerappendix}

Our objective is to find the dynamics with the effective Hamiltonians 
\begin{dmath*}
\bar{H}_1=-\lambda_{1}\sum_{j=1}^{N-1}\sigma_{y}^{(j)}\sigma_{y}^{(j+1)},
\end{dmath*}
while 
\begin{dmath*}
\bar{H}_2 = \sum_{j=1}^{N-1}\left\lbrace\frac{(\lambda_{1}+\lambda_{2})}{2}\left[\sigma_{x}^{(j)}\sigma_{y}^{(j+1)}+\sigma_{y}^{(j)}\sigma_{x}^{(j+1)}\right]-\lambda_{1}\left[\sigma_{y}^{(j)}\sigma_{y}^{(j+1)}\right]\right\rbrace.
\end{dmath*}
As mentioned before, we largely follow the approach presented in Refs.~\cite{LiebMattisarticle,AmicoJPA2004}. First, we introduce the raising and lowering operators $a_i^\dagger = \frac{1}{2}(\sigma_x^{(i)} + i\sigma_y^{(i)})$ and $a_i = \frac{1}{2}(\sigma_x^{(i)} - i\sigma_y^{(i)})$. Thereafter, the fermionic operators $c_i$ and $c_i^\dagger$ are defined as 
\begin{align*}
c_i = \exp\left[\pi i \sum_{j=1}^{i - 1} a_j^\dagger a_j \right] a_i, \\
c_i^\dagger = a_i^{\dagger}\exp\left[-\pi i \sum_{j=1}^{i - 1} a_j^\dagger a_j \right]. \\
\end{align*}
Using the Jordan-Wigner transformation for $\bar{H}_{2}$, we get
\begin{dmath}
\label{H2JW}
\bar{H}_{2}=\lambda_{1}\sum_{i,j}\left[c_{i}^{\dagger}J_{ij}c_{j}+\frac{1}{2}(e^{-i\phi}c_{i}^{\dagger}K_{ij}c_{j}^{\dagger}+ \text{h.c.})\right],
\end{dmath}
where

\[
J=-
  \begin{pmatrix}
    0 & 1 & 0 &  &   &   &  &   \\
    1 & 0 & 1 & 0 &   &   &  &  \\
    0 & 1 & 0 & 1 & 0 &  &  & \\
      & . & . & . & . & . &. & \\
      &   & . & . & . & . & . & \\
          &   &  & . & . & . & . & . \\
      &   &   &   & 0 & 1 & 0 &1 \\
      &   &   &  &   & 0 & 1 & 0 \\
  \end{pmatrix}
,\]

\[
K=\gamma
  \begin{pmatrix}
    0 & 1 & 0 &  &   &   &  &   \\
    -1 & 0 & 1 & 0 &   &   &  &  \\
    0 & -1 & 0 & 1 & 0 &  &  & \\
      & . & . & . & . & . &. & \\
      &   & . & . & . & . & . & \\
          &   &  & . & . & . & . & . \\
      &   &   &   & 0 & -1 & 0 &1 \\
      &   &   &  &   & 0 & -1 & 0 \\
  \end{pmatrix}
,\]

and
\begin{dmath*}
 \gamma=\frac{\sqrt{\lambda_{1}^2+(\lambda_{1}+\lambda_{2})^2}}{\lambda_{1}},
 \end{dmath*}
\begin{dmath*}
\phi=\arctan{[(\lambda_{1}+\lambda_{2})/\lambda_{1}]}.
\end{dmath*}
The form of $\bar{H}_1$ after the Jordan-Wigner transformation is found to be the same as that in Eq.~\eqref{H2JW}, except that now $\phi = 0$ and $\gamma = 1$. 
 
Following Ref.~\cite{LiebMattisarticle}, we now look for a linear transformation of the form 
\begin{dmath}
\eta_{k}=\sum_{i} g_{ki}c_{i}+h_{ki}c_{i}^{\dagger},
\label{lineartransformation}
\end{dmath}
\begin{dmath}
\eta_{k}^{\dagger}=\sum_{i} g_{ki}^{*}c_{i}^{\dagger}+h_{ki}^{*}c_{i},
\label{lineartransformation2}
\end{dmath}
such that $\bar{H}_2$ becomes
\begin{dmath}
\label{H2bardiagonal}
\bar{H}_2=\sum_{k}\Lambda_{k}\eta_{k}^{\dagger} \eta_{k}\, + \, \text{constant}.
\end{dmath}
The constant term is unimportant. Our central task in finding the dynamics is to find the eigenvalues $\Lambda_k$. To do this, we note that if Eq.~\eqref{H2bardiagonal} is true, then
\begin{dmath}
\label{condition}
[\eta_{k}, H]-\Lambda_{k}\eta_{k}=0
\end{dmath}
Substituting Eq.~\eqref{lineartransformation} in Eq.~\eqref{condition}, we get

\begin{dmath}
\sum_{m} [g_{km}J_{mj}-e^{i\phi}h_{km}K_{mj}] = \frac{\Lambda_{k}}{\lambda_{1}}g_{kj},
\label{coupled1}
\end{dmath}

\begin{dmath}
\sum_{m} [g_{km}K_{mj}e^{-i\phi}-h_{km}J_{mj}] = \frac{\Lambda_{k}}{\lambda_{1}}h_{kj}.
\label{coupled2}
\end{dmath}
These are further simplified by introducing the linear combinations
\begin{dmath*}
\Phi_{ki}=g_{ki}+e^{i\phi}h_{ki},
\end{dmath*}
\begin{dmath*}
\Psi_{ki}=g_{ki}-e^{i\phi}h_{ki}.
\end{dmath*}
Eqs.~\eqref{coupled1} and \eqref{coupled2} can be cast in the form of matrix equations as 
\begin{dmath}
\label{coupled3}
\Phi_{k}(J-K)=\frac{\Lambda_{k}}{\lambda_{1}}\Psi_{k},
\end{dmath}
\begin{dmath}
\label{coupled4}
\Psi_{k}(J+K)=\frac{\Lambda_{k}}{\lambda_{1}}\Phi_{k},
\end{dmath}
 in where $\Psi_{k}$ and $\Phi_{k}$ denote the $k^{\text{th}}$ row of matrices $\Phi$ (whose matrix elements are given by $\Phi_{ki}$) and $\Psi$ (whose matrix elements are given by $\Psi_{ki}$) respectively.
Eliminating $\Psi_{k}$, we get  
\begin{dmath}
\label{decoupled}
\Phi_{k}(J-K)(J+K)=\left(\frac{\Lambda_{k}}{\lambda_{1}}\right)^{2}\Phi_{k}.
\end{dmath}
We then view Eq.~\eqref{decoupled} as an eigenvalue problem to solve for $\Lambda_{k}$. However, for $\bar{H}_1$, it turns out that $\Lambda_{k}$ can be zero, therefore $\Phi_{k}$ and $\Psi_{k}$ are solved using Eqs.~\eqref{coupled1} and \eqref{coupled2} as a null space problem. 

We now aim to find the concurrence for any two spins in our spin chain. Since the Pauli matrices form a basis, we can write the two-spin state as 
\begin{dmath*}
\rho^{(ij)}(t) = \text{Tr}_{ij}[\rho_{\text{tot}}(t)]=\frac{1}{4}\sum_{\alpha \beta} \Theta_{\alpha \beta}^{i,j}(t) \sigma_{\alpha}^{(i)}\otimes \sigma_{\beta}^{(j)},
\end{dmath*}
where we have introduced the time-dependent correlation functions $\Theta_{\alpha \beta}^{i,j} = \langle \sigma_\alpha^{(i)} \sigma_\beta^{(j)} \rangle = \text{Tr}_{ij} [\rho^{(ij)}(t) \sigma_\alpha^{(i)} \sigma_\beta^{(j)}]$, and $\rho_{\text{tot}}(t)$ is the density matrix for the complete spin chain. Once we can figure out these correlation functions, we have the relevant two-spin density matrix, and thereby the concurrence. To calculate the correlation functions, we push the time dependence to the operators. We define 
\begin{dmath*}
c_{i}(t)=e^{i\bar{H}_{2}t} c_{i}e^{-i\bar{H_{2}}t},
\end{dmath*}
\begin{dmath*}
c_{i}^\dagger(t)=e^{i\bar{H}_{2}t} c_{i}^\dagger e^{-i\bar{H_{2}}t}.
\end{dmath*}
The dynamics for $\bar{H}_1$ can be found analogously. To find the time-evolving operators $c_i(t)$ and $c_i^\dagger(t)$, the strategy is to first transform the operators $c_i$ and $c_i^\dagger$ to the operators $\eta_k$ and $\eta_k^\dagger$, since the Hamiltonian is diagonal in terms of $\eta_k$ and $\eta_k^\dagger$. We then transform back to the operators $c_i$ and $c_i^\dagger$. The result is that we can write 
\begin{align*}
c_{i}(t) = \sum_l A_{il}(t)c_{l}+B_{il}(t)c_{l}^{\dagger},
\end{align*}
and
\begin{align*}
c_{i}^{\dagger}(t )= \sum_l A_{il}^{*}(t)c_{l}^{\dagger}+B_{il}^{*}(t)c_{l},
\end{align*}
where the matrices $A$ and $B$ are defined as 
\begin{align*}
A(t)=g'(t)g + h'(t)h^{*},\\
B(t)=g'(t)h+h'(t)g^{*}.
\end{align*}
Here $g'(t)=g' e^{-it\mathcal{H}}$ and $h'(t)=h' e^{it\mathcal{H}}$, with
\[
\mathcal{H}=\lambda_1
  \begin{pmatrix}
    \Lambda_{1} & . & . &  &   &   &  &   \\
    0 &  \Lambda_{2} &  & . &   &   &  &  \\
    0 & 0 & \Lambda_{3} &  & 0 &  &  & \\
      & . & . & . & . & . &. & \\
      &   & . & . & . & . & . & \\
          &   &  & . & . & . & . & . \\
      &   &   &   & 0 & 0 &  \Lambda_{N-1} &0 \\
      &   &   &  &   & 0 & 0 &  \Lambda_{N} \\
  \end{pmatrix}
.\]
The matrices $g$ and $h$ are the transformation matrices given in Eqs.~\eqref{lineartransformation} and \eqref{lineartransformation2}, while $g'$ and $h'$ are the inverse transformation matrices, that is, 
\begin{dmath*}
c_{i}= \sum_j g'_{ij}\eta_{j}+h'_{ij}\eta_{j}^{\dagger},
\end{dmath*}

\begin{dmath*}
c_{i}^\dagger = \sum_j g^{'*}_{ij}\eta_{j}^\dagger + h_{ij}^{'*}\eta_{j}.
\end{dmath*}
With the matrices $A$ and $B$ at hand, we calculate the correlation functions. For example,
\begin{dmath}
    \Theta_{xx}^{l, m}(t)= \bra{\psi}Q_{l}(t)P_{l+1}(t)Q_{l+1}(t)...P_{m-1}(t)Q_{m-1}(t)P_{m}(t)\ket{\psi},
\end{dmath}
where $P_{l}(t)=c_l^\dagger(t) + c_l(t)$ and $Q_{l}(t)=c_l^\dagger(t) - c_l(t)$. We now choose our spin chain state to be $\ket{11\hdots1}$. Just like the results in Refs.~\cite{McCoyXY2,AmicoJPA2004} for the standard XY model, $\Theta_{xx}^{l,m}(t)$ can be expressed in Pfaffian form, that is,
\vspace{1cm}
\begin{widetext} 
\begin{align}
\Theta_{xx}^{l, m}(t)= \text{pf}\begin{pmatrix}
    0 & F_{l,l+1} & S_{l,l+1} & \cdots & S_{l, m-1} & F_{l,m}\\
    &  0 & W_{l+1, l+1}&\cdots& W_{l+1, m-1}&T_{l+1, m}\\
    &  &  &\cdots & \cdot &\cdot&\\
      & & & &W_{m-1, m-1}&T_{m-1, m}\\
      &   &  &  &  &F_{m-1, m}\\
          &   &  & &  &0 \\
  \end{pmatrix},
\end{align}

Similarly, we also obtain

\begin{align}
\Theta_{yy}^{l, m}=(-1)^{m-l} \text{pf}\begin{pmatrix}
    0 & W_{l,l+1} & T_{l,l+1} & \cdots & T_{l, m-1} & W_{l,m}\\
    &  0 & F_{l+1, l+1}&\cdots& F_{l+1, m-1}&S_{l+1, m}\\
    &  &  &\cdots & \cdot &\cdot&\\
      & & & &F_{m-1, m-1}&S_{m-1, m}\\
      &   &  &  &  &W_{m-1, m}\\
          &   &  & &  &0 \\
  \end{pmatrix},
\end{align}

\begin{align}
\Theta_{xy}^{l, m}(t)=-i \, \text{pf}\begin{pmatrix}
    0 & F_{l,l+1} & S_{l,l+1} & \cdots & S_{l, m-1} & S_{l,m}\\
    &  0 & W_{l+1, l+1}&\cdots& W_{l+1, m-1}&W_{l+1, m}\\
    &  &  &\cdots & \cdot &\cdot&\\
      & & & &W_{m-1, m-1}&W_{m-1, m}\\
      &   &  &  &  &S_{m-1, m}\\
          &   &  & &  &0 \\
  \end{pmatrix},
\end{align}

\begin{align}
\Theta_{yx}^{l, m}(t)=-i \, \text{pf}\begin{pmatrix}
    0 & T_{l,l+1} & W_{l,l+1} & \cdots & W_{l, m-1} & T_{l,m}\\
    &  0 & W_{l+1, l+1}&\cdots& W_{l+1, m-1}&T_{l+1, m}\\
    &  &  &\cdots & \cdot &\cdot&\\
      & & & &W_{m-1, m-1}&T_{m-1, m}\\
      &   &  &  &  &F_{m-1, m}\\
          &   &  & &  &0 \\
  \end{pmatrix},
\end{align}

\begin{align}
\Theta_{zz}^{l, m}(t)= \text{pf}\begin{pmatrix}
    0 & W_{l,l} & T_{l,m} & W_{l, m}\\
    &  0 & F_{l,m}&S_{l, m}\\
    &  &0& W_{m,m}\\
      & & &0&\\
  \end{pmatrix},
\end{align}
\end{widetext}
where
\begin{align*}
F_{i, j}&=\exs{Q_{i}(t)P_{j}(t)},\\
S_{i, j}&=\exs{Q_{i}(t)Q_{j}(t)},\\
T_{i, j}&=\exs{P_{i}(t)P_{j}(t)},\\ 
W_{i, j}&=\exs{P_{i}(t)Q_{j}(t)}.
\end{align*}
Since the initial state is $\ket{11\hdots1}$,
\begin{align}
 \langle Q_{i}(t)P_{j}(t) \rangle &= (Y(t)X^{\dagger}(t))_{i, j}, \\
 \langle Q_{i}(t)Q_{j}(t) \rangle &= -(Y(t)Y^{\dagger}(t))_{i, j}, \\
 \langle P_{i}(t)P_{j}(t) \rangle &= (X(t)X^{\dagger}(t))_{i, j}, \\
 \langle P_{i}(t)Q_{j}(t)\rangle &= -(X(t)Y^{\dagger}(t))_{i, j}.
\end{align}
Here $X(t)$ and $Y(t)$ are calculated in terms of $A(t)$ and $B(t)$
as
\begin{align*}
 X(t)=A(t)+B(t)^{*}\\
Y(t)=B(t)^{*}-A(t)
\end{align*}
We also find that $\Theta_{z, 0}^{l, m}(t)= \langle \sigma_z^{(l)} \rangle = -W_{l,l}$. All the other correlation functions are zero \cite{AmicoJPA2004}. With the correlation functions now known, we can find the two-spin density matrix and hence the concurrence.

\section{What if $n_y$ is not exactly equal to $2n_x$?}

\label{appendixny2nx}

\begin{figure}[b]
\centering
\includegraphics[width=.8\linewidth]{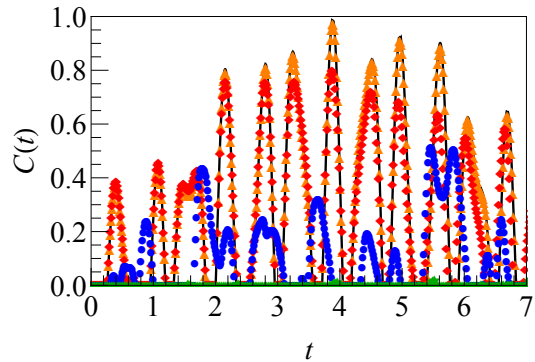}
\caption{(Color online) Dynamics of the entanglement between spins $1$ and $4$ for the XYZ model with $N = 4$. We have used $\lambda_1 = 2$, $\lambda_2 = 1$, and $\lambda_3 = -5$ (same as Fig.~\ref{Hbar1Hbar2XYZspecial}). We have set $n_x = 1$, and vary the value of $n_y$. We show the dynamics with $n_y = 2.00001$ (orange triangles), $n_y = 2.0001$ (red diamonds), $n_y = 2.01$ (blue circles), and $n_y = 2.1$ (green stars). The solid, black line shows the dynamics with the effective Hamiltonian $\bar{H}_2$, while $\bar{H}_1$ (not shown) leads to no entanglement. The green stars largely overlap with the horizontal axis. The initial state is $\ket{1111}$.}
\label{fignynxrobustness}
\end{figure}

We have shown that choosing special control fields such that $n_y = 2n_x$ can lead to better performance. A natural question that then arises is to investigate how stringently this condition needs to be met. To check this, we have considered a spin chain $N = 4$ and numerically solved the Schrodinger equation in the presence of the control fields with $n_y$ not necessarily equal to $2n_x$ to see how closely the dynamics generated by $\bar{H}_2$ are reproduced. As illustrated in Fig.~\ref{fignynxrobustness}, we have found that as $n_y$ approaches $2n_x$, the dynamics with the full time-dependent Hamiltonian approach the dynamics with the effective Hamiltonian $\bar{H}_2$. Moreover, $n_y$ needs to very close to $2n_x$ for the exact dynamics to be effectively the same as those generated by $\bar{H}_2$. That is, if the frequencies used are in the GHz regime [see Fig.~\ref{statetransferN4} caption], then the error in the frequencies needs to be less than around $10$ kHz. However, even if the condition $n_y = 2n_x$ is not met so stringently, the entanglement generated can be significant as illustrated by the red diamonds and the blue circles in Fig.~\ref{fignynxrobustness}. We obtain similar results for $N = 8$.

\end{document}